\documentclass[12pt,letterpaper]{article}
\usepackage{epsfig,rotating,setspace,latexsym,amsmath,epsf,amssymb,amsfonts,bm,theorem,cite,caption,subcaption,enumerate,longtable,accents,graphicx,authblk,epstopdf,url,color,multirow,mathtools,comment,tabularx,physics,algorithm,dsfont}
\usepackage{algpseudocode}
\usepackage[short,c2]{optidef}
\usepackage[toc,title]{appendix}

\DeclareMathOperator*{\argmax}{arg\,max}

\newtheorem{theorem}{Theorem}

\newtheorem{corollary}{Corollary}
\newtheorem{definition}{Definition}
\newtheorem{remark}{Remark}
\newtheorem{lemma}{Lemma}

\newcolumntype{Y}{>{\centering\arraybackslash}X}
\newcommand{\e}{{\mathbb{E}}}

\allowdisplaybreaks

\title{Beyond Martingale Estimators: Structured Estimators for Maximizing Information Freshness in \\ Query-Based Update Systems\thanks{This work is presented in part at the IEEE Information Theory Workshop, September 2025.}}

\author[1]{Sahan Liyanaarachchi}
\author[1]{Sennur Ulukus}
\author[2]{Nail Akar}

\affil[1]{\normalsize University of Maryland, College Park, MD, USA}
\affil[2]{\normalsize Bilkent University, Ankara, T\"{u}rkiye}

\begin{document}
\date{}
\maketitle

\vspace*{-1.0cm}

\begin{abstract}
    This paper investigates information freshness in a remote estimation system in which the remote information source is a continuous-time Markov chain (CTMC). For such systems, estimators have been mainly restricted to the class of martingale estimators in which the remote estimate at any time is equal to the value of the most recently received update. This is mainly due to the simplicity and ease of analysis of martingale estimators, which however are far from optimal, especially in query-based (i.e., pull-based) update systems. In such systems, maximum a-posteriori probability (MAP) estimators are optimal. However, MAP estimators can be challenging to analyze in continuous-time settings. In this paper, we introduce a new class of estimators, called \emph{structured estimators}, which can seamlessly shift from a martingale estimator to a MAP estimator, enabling them to retain useful characteristics of the MAP estimate, while still being analytically tractable. Particularly, we introduce a new estimator termed as the $p$-MAP estimator which is a piecewise-constant approximation of the MAP estimator with finitely many discontinuities, bringing us closer to a full characterization of MAP estimators when modeling information freshness. In fact, we show that for time-reversible CTMCs, the MAP estimator reduces to a $p$-MAP estimator. Using the {\em binary freshness} (BF) process for the characterization of information freshness, we derive the freshness expressions and provide optimal state-dependent sampling policies (i.e., querying policies) for maximizing the mean BF (MBF) for pull-based remote estimation of a single CTMC information source, when structured estimators are used. Moreover, we provide optimal query rate allocation policies when a monitor pulls information from multiple heterogeneous CTMCs with a constraint on the overall query rate. We present numerical examples to validate the proposed approach.
\end{abstract}

\section{Introduction} \label{sec:intro}
With the dawn of 6G communications, spanning from modern day vehicular networks to the outskirts of cislunar communications, timeliness of information has become a crucial feature that must be integrated into any contemporary communication infrastructure \cite{AoI_self_drive, yuan_towards,sahan-cislunar}. Henceforth, a variety of metrics, such as, age of information (AoI) \cite{yates2020age, age1, age2, rts2012, rtsms2012, lts2015, ys17, iot_aoi_2019, age_constrained_transmission, early_sample, cyclic_scheduler_design, shared_servers}, age of incorrect information (AoII) \cite{AoII2019, AoII_Markov, AoIV}, version age of information (VAoI) \cite{yates21gossip, Abolhassani21version, melih2020infocom} and binary freshness (BF) \cite{melih_BF_cache, melih_IF_CUS, melih_BF_Inf, melih_BF_gossip} have been proposed to evaluate and quantify the timeliness of information in such freshness-critical systems. Among them, BF has been the most widely used metric in the characterization of information freshness when monitoring continuous-time Markov chain (CTMC) information sources, due to its direct relation to error probability.

Several variations of BF have been proposed in the literature such as fresh when equal (FWE), fresh when close (FWC) and fresh when sampled (FWS), which incorporate various semantic aspects of the system \cite{nail_QS}. In this work, we adopt the traditional FWE variation of BF. To formally define the metric, let us denote by $\Delta(t)$, the BF process at time $t$, where $\Delta(t)=1$, if the estimator is in sync with the source, and zero otherwise. When $\Delta(t)$ is stationary and ergodic, the mean binary freshness (MBF), denoted by $\e[\Delta]$, can be expressed as the following time average,
\begin{align}
    \e[\Delta]=\limsup_{T\to\infty}\frac{1}{T}\e\left[\int_0^T\mathds{1}\{X(t)=\hat{X}(t)\} \dd{t} \right],
\end{align}
where $\Delta$ is the steady-state random variable associated with the marginal distribution of the random process $\Delta(t)$, $\mathds{1}\{\cdot\}$ is the indicator function, $X(t)$ is the stochastic process that is being monitored, and $\hat{X}(t)$ is its estimate used to track the process. In general, larger values of MBF (which ranges between 0 and 1) are indicative of good tracking performance. 

Majority of the prior works that considers BF is restricted to the analysis of BF under the martingale estimator (ME) \cite{nail_QS}. This is mainly due to the simplicity of the ME, which uses the most recently received sample as its current estimate of the remote process. However, ME can be detrimental from a freshness perspective, especially in pull-based update systems \cite{Goal_oriented_pull_based, pragmatic_com, push_pull_goal_com,push_pull_medium_access}. For example, under the event that we have sampled a less probable state, the ME will be forced to retain this less probable state as its estimate, until the next sampling instance. For such systems, MAP estimators are known to be ideal, but they come with their own set of challenges. For instance, MAP estimators can be unstable or infinitely oscillating (see Section~\ref{sec:prelim}), and hence obtaining closed-form expressions for MBF can be cumbersome. Motivated by this, in this paper, we introduce a new class of estimators, which we term as {\em structured estimators}, that can seamlessly shift between the martingale and MAP estimators. As the name suggests, these estimators evolve based on a predetermined structure during inter-update periods. Among these structured estimators, we introduce a specific estimator termed as the $p$-MAP estimator, which approximates the MAP estimator with a piecewise-constant function with finitely many transition points (finite number of regions where the function changes). We show that the $p$-MAP estimator coincides with the MAP estimator for time-reversible CTMCs (see Section~\ref{sec:prelim}) if the transition points are chosen appropriately. Even for infinitely oscillating MAP estimators, the $p$-MAP estimator can provide a close enough approximation as we increase the number of transition points, i.e., intermediate stages.

\begin{figure}[t]
    \centering
    \includegraphics[width=0.9\textwidth]{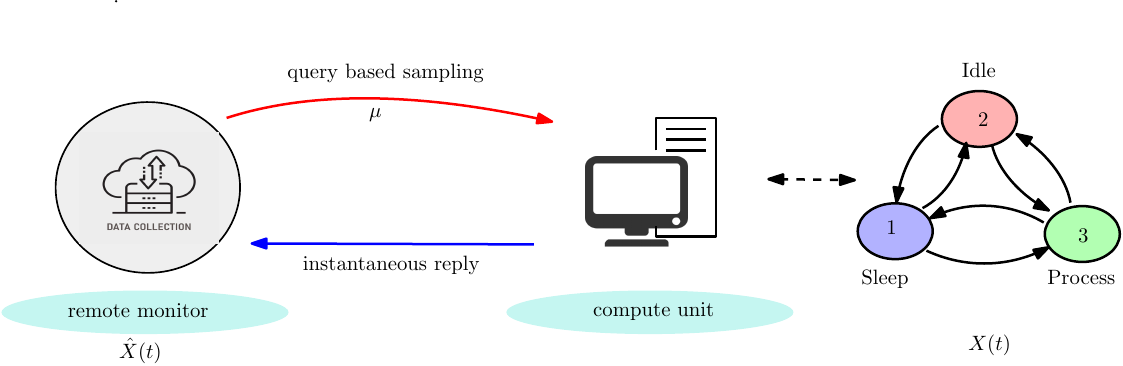}
    \caption{Query-based sampling of CTMCs. In this example, the remote CTMC represents the state of a compute unit, which takes three values $\{\text{Sleep}, \text{Idle}, \text{Process}\}$. The true state of the CTMC at time $t$ is $X(t)$. The remote monitor samples (sends queries) to the CTMC, and gets instantaneous responses (readings of the state). Based on the responses, the monitor keeps an estimate of the state of the remote CTMC at time $t$ as $\hat{X}(t)$.}
    \label{fig:sys_model}
\end{figure}

\begin{figure}[t]
    \centering
    \begin{subfigure}[b]{0.45\textwidth}
        \centering
        \includegraphics[width=0.9\textwidth]{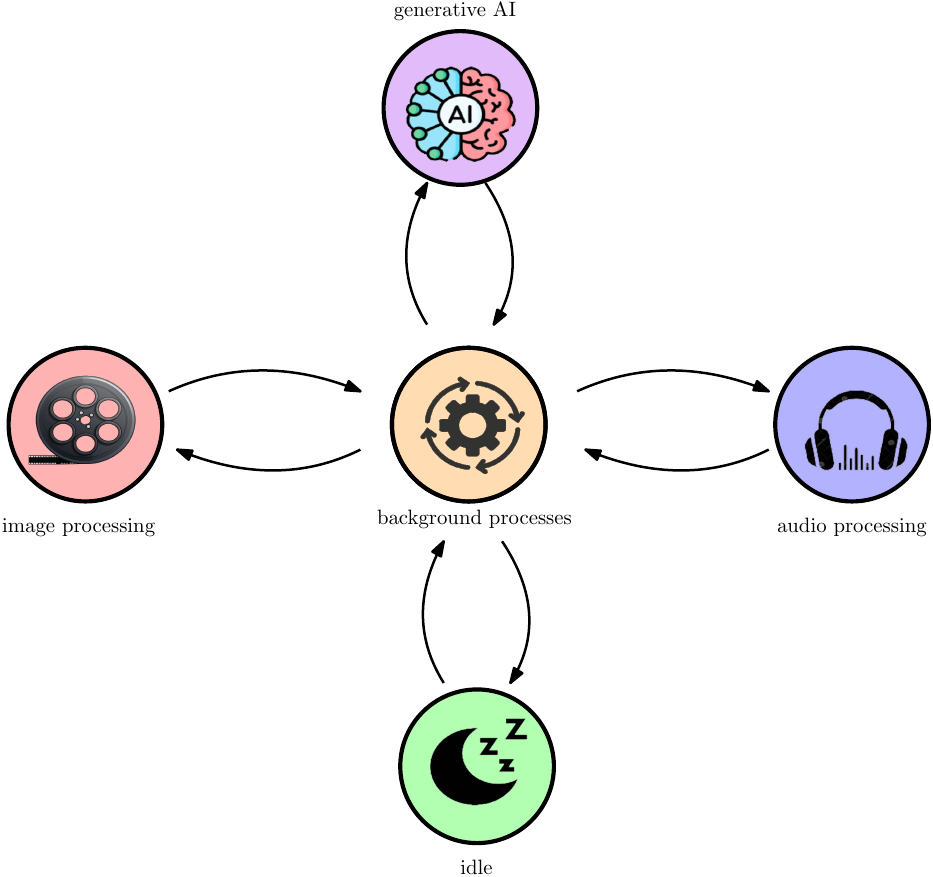}
        \caption{Task scheduling process in CPUs.}
        \label{fig:task_scheduling}
    \end{subfigure}
    \hfill 
    \begin{subfigure}[b]{0.45\textwidth}
        \centering
        \includegraphics[width=\textwidth]{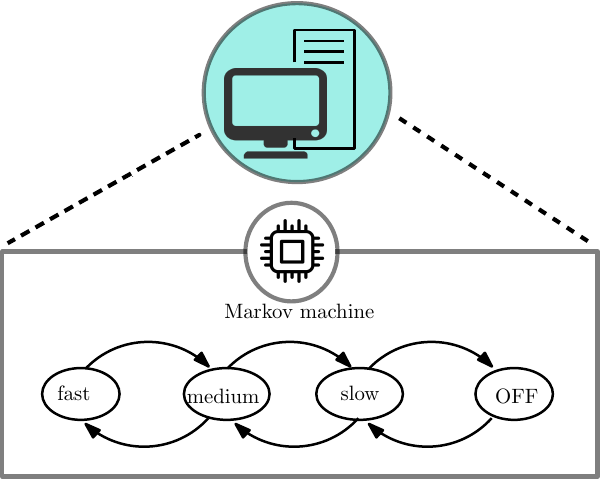}
        \caption{Sequential Markov machine.}
        \label{fig:seq_mm}
    \end{subfigure}
    \caption{Examples of time-reversible CTMC structures.}
    \label{fig:tr_examples}
\end{figure}

In this paper, we study the proposed structured estimators in the context of a remote estimation system which employs a query-based sampling procedure to monitor a CTMC information source, where a remote monitor inquires about the state of the CTMC by sending queries to the CTMC with exponentially distributed inter-query times at a fixed rate (see Fig.~\ref{fig:sys_model}). Upon receiving a query, the source instantaneously sends back its status to the monitor. We assume that query and status transmission times are negligible when compared to the sojourn times of the CTMC source, and the study of non-zero transmission times is left for future research. Under this setting, we provide closed-form expressions for MBF when using the proposed structured estimators. In particular, we dedicate a significant portion of our study to derive the exact closed-form expressions for MBF for time-reversible CTMCs. However, the main theorems and techniques we have developed can also be applied  to  more general CTMCs that may not necessarily be time-reversible. The particular interest in time-reversible  CTMCs is two fold. First, such CTMCs give rise to stable MAP estimators, where stable in this context refers to estimators whose values do not oscillate or change indefinitely (they converge to a fixed value after a finite amount of time). This enables us to find closed-form expressions easily and thereby relax the computational burden of some of our algorithms. Second, they are the ones that typically arise in remote estimation frameworks. For example, suppose we want to monitor a compute unit that frequently shifts between its main program and its subroutines (see Fig.~\ref{fig:tr_examples}(a)) or a birth and death chain (BDC) which can model sequential Markov machines or queuing systems (see Fig.~\ref{fig:tr_examples}(b)). These structures are all reminiscent of time-reversible CTMCs.

We also look into the problem of monitoring multiple heterogeneous CTMCs under our structured estimators and find the optimal sampling rate that should be allocated to each CTMC when constrained by an overall sampling budget. The optimization framework used for this problem is not restricted to time-reversible CTMCs and is applicable for monitoring any general CTMC which is not necessarily time-reversible. In addition to the derivation of exact MBF expressions for query-based sampling with fixed query rates, we  also study state-dependent (i.e., observation-dependent) sampling policies in which the query rate (i.e., the sampling rate) is allowed to depend on the most recently received update. In particular, we provide a semi-Markov decision process (SMDP) formulation along with a policy iteration algorithm to find the optimal state-dependent sampling policy for general CTMCs which are not necessarily time-reversible. We show through numerical examples that state-dependent policies improve MBF significantly. Through our analytical expressions and extensive numerical experiments, we show that significant gains are attainable in terms of MBF with the use of the proposed structured estimators over conventional ME. Moreover, we show that these gains can further be improved by employing a state-dependent sampling policy.

Our contributions are summarized below:
\begin{itemize}
    \item We extend the query-based sampling framework for CTMCs with fixed query rate and ME, proposed in \cite{nail_QS}, to a wider class of estimators, where we show that we can obtain better freshness figures by allowing the estimator to be updated during inter-sampling periods, which is not possible with an ME. 
    \item We extend this general class of estimators to the monitoring of multiple CTMCs similar to \cite{nail_QS}, which studied the same problem under an ME.
    \item We propose state-dependent sampling policies when monitoring a CTMC. This was not explored in \cite{nail_QS}, and adds a new dimension to the query-based sampling problem. We obtain the optimal query-based sampling strategy using the SMDP formulation, and we show that we can further improve the MBF by combining state-dependent sampling with the proposed structured estimators. 
\end{itemize}

The remainder of our paper is organized as follows. Section~\ref{sec:rel_works} presents the related work on this topic. Section~\ref{sec:prelim} provides a theoretical overview of CTMCs that would be helpful to present the motivation and the construction of our estimators. Section~\ref{sec:sys_model} describes our system model and the various structured estimators considered in this work. In Section~\ref{sec:main_results}, we present our main analytical results on MBF for these structured estimators. In Section~\ref{sec:state_dep}, we present the SMDP framework to find the optimal state-dependent sampling policy and in Section~\ref{sec:mult_ctmcs}, we present techniques to find the optimal rate allocation policy while monitoring multiple CTMCs. Finally, we verify our theoretical findings through simulations in Section~\ref{sec:numer}.

\section{Related Work} \label{sec:rel_works}
The work in \cite{AoI_Markov} studies the problem of remote estimation of a discrete-time Markovian source, where they show that a sampling policy that minimizes the AoI does not necessarily minimize the probability of error. This work was then extended to the AoII metric in \cite{AoII_Markov}, where they study the problem of minimizing the AoII for a binary discrete-time Markov source. There, the authors restrict their analysis to binary symmetric Markov chains with inter-state transition probability chosen as $p<\frac{1}{2}$. The main reason for this restriction is that, when $p<\frac{1}{2}$, the ME coincides with the MAP estimator, making the analysis simpler. Another avenue in this domain, which considers a MAP estimator, is the work done in \cite{ismail_map}. Here, the authors consider the problem of AoII minimization with the use of a MAP estimator, for discrete-time Markov sources. However, they resort to reinforcement learning (RL) techniques to circumnavigate some of the challenges posed by the MAP estimate. 

On a similar note, the work in \cite{real_time_tracking_POMDP} also considers a non-martingale estimator where they look into the problem of monitoring a binary Markov source over an error-prone channel using two heterogeneous sensors. They formulate the problem as a partially observable Markov decision process (POMDP) and find the optimal scheduling strategy to minimize a distortion function. Here, they choose the estimator so as to minimize the expected distortion given all the information available. However, their work is restricted to the discrete domain and a single binary Markov source. The work in \cite{imperfect_feedback_deep_RL} also studies a problem that shares similar characteristics to the above. They look into the problem of minimizing a generic distortion function when monitoring a finite-state symmetric Markov source in an energy harvesting setting. Similar to \cite{real_time_tracking_POMDP}, they also formulate the problem as a POMDP and utilize deep reinforcement learning (DRL) techniques to solve it. The works in \cite{Goal_oriented_pull_based} and \cite{semantic_aware_POMDP} are a few other works that follow the same approach to tackle problems in the same domain.

The problem of query-based sampling of CTMCs was introduced in \cite{nail_QS}, where the authors studied the problem of maximizing different variants of MBF when monitoring multiple heterogeneous CTMCs. However, in \cite{nail_QS}, ME was considered as the estimator for simplicity. In our work, we solve the same problem under a more general estimator yielding better MBF results. On a similar note, the work in \cite{melih_BF_Inf} looks into the problem of utilizing the MBF metric to track an infected population during pandemics such as Covid-19. They model the infected individuals as binary CTMCs and find the optimal rates at which individuals should be tested (i.e., their states sampled) by modeling the testing times as exponentially distributed random variables. For this problem, they deviate from the traditional MBF definition, and utilize a weighted freshness metric penalizing false positives and false negatives, differently. In \cite{melih_BF_Inf} also, the analysis is again carried out using the ME.

The work in \cite{Markov_machines} looks into the problem of query-based sampling of Markov machines, where they consider the problem of maximizing MBF and other related metrics. However, they too resort to an ME which may lead to a crude decision process. This work was later extended in \cite{revMax}, where the authors used the MAP estimator of the system to improve the decision process so as to maximize the revenue. Another similar avenue is the work in \cite{graves2024}, which looks into the problem of monitoring multiple worker nodes in a distributed computation setting, where each worker is modeled as a binary discrete-time source with free and busy states. In \cite{graves2024}, each source is assigned a different priority level and the authors find the optimal update pattern that should be followed by these sources so as to minimize the error probability of the top-$k$ set of these sources. However, here also, the estimator is restricted to be an ME. 

All of these works look into important applications where the MBF metric is admissible. However, most of these works rely on the simplicity of the ME when evaluating the MBF metric. In our work, we take upon the challenge of assessing the MBF metric and finding optimal rate allocation schemes under more structurally sophisticated estimators that yield better freshness, and thereby lower error probability.

\section{Preliminaries} \label{sec:prelim}
Let $X(t)$ be a finite-state irreducible continuous-time Markov chain (CTMC) with $S$ states in $\mathcal{S}=\{1,2,\dots,S\}$, where $X(t)\in \mathcal{S}$. Let $Q=\{q_{ij}\}_{i,j\in \mathcal{S}}$ be the generator matrix of this CTMC, where $q_{ii}=-\sum_{j\neq i}q_{ij}=-q_i$. Therefore, $Q\bm{1}=\bm{0}$, where $\bm{1}$ is a column vector of all ones, $\bm{0}$ is a column vector of all zeros. Since $X(t)$ is irreducible and has a finite state space, its stationary distribution denoted by the column vector $\bm{\pi}=\{\pi_1,\pi_2,\dots,\pi_S\}$ exists and satisfies $\bm{\pi}^TQ=\bm{0}$. Moreover, the transition rate matrix $P(t)=e^{Qt}$ of the CTMC will converge to $\mathbf{1}\bm{\pi}^T$ as $t\to \infty$ (ergodic). Hence, we have the following intriguing property given in Lemma~\ref{lem:T} as a consequence of the ergodicity of the CTMC.

\begin{lemma}\label{lem:T}
    If $i^*=\argmax_{i\in \mathcal{S}}\pi_i$ is unique, then $\exists~\tau^*<\infty$, such that $\argmax_{i \in \mathcal{S}} \bm{e}_j^T P(t)=i^*$, for all $t>\tau^*$ and for all $j\in \mathcal{S}$, where $\bm{e}_j$ is a vector of all zeros except for the $j$th entry which is one.
\end{lemma}

\begin{figure}[t]
    \centering
    \begin{subfigure}[b]{0.45\textwidth}
        \centering
        \includegraphics[width=0.8\textwidth]{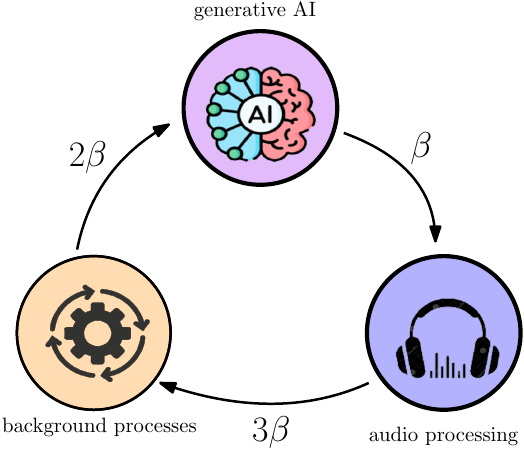}
        \caption{Non time-reversible CTMC.}
        \label{fig:unique_max_ctmc}
    \end{subfigure}
    \hfill 
    \begin{subfigure}[b]{0.45\textwidth}
        \centering
        \includegraphics[width=\textwidth]{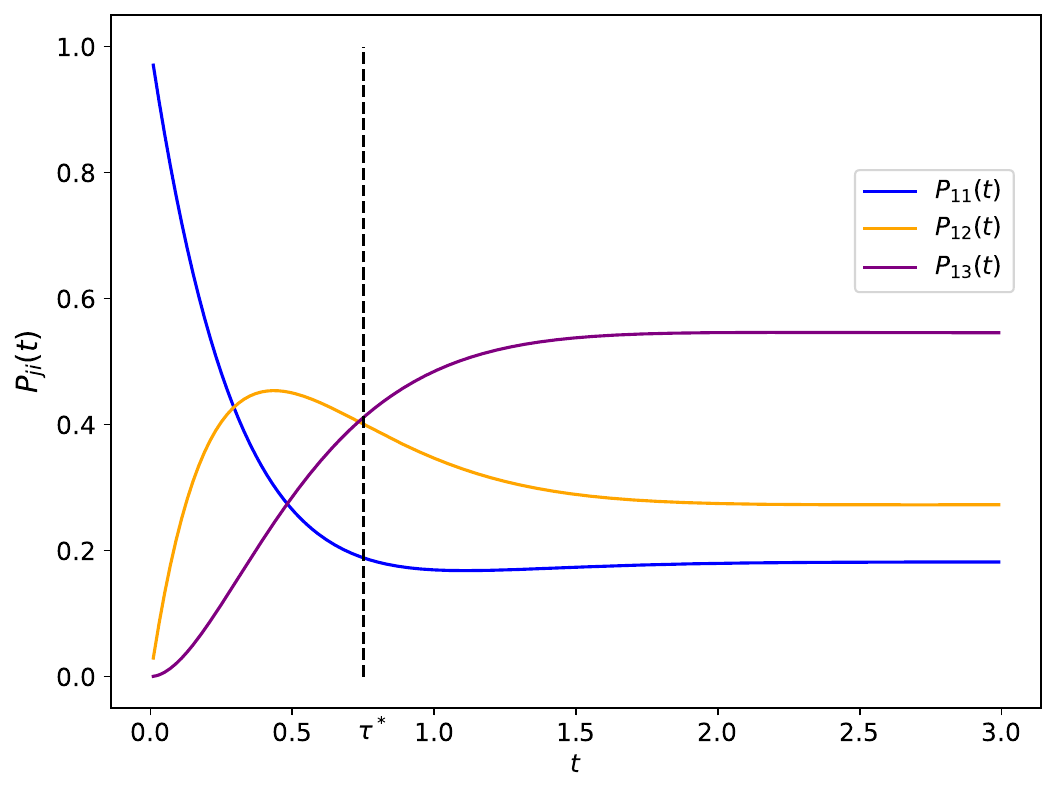}
        \caption{Transition probabilities.}
        \label{fig:osc_tmap}
    \end{subfigure}
    \caption{(a) A CTMC with a unique stationary maximum. Here, the stationary distribution is $\bm\pi=\{0.18,0.27,0.55\}$, and hence, $i^*=\argmax_{i\in \mathcal{S}}\pi_i=3$. (b) Note that, for all $t$ after $\tau^*$, we have $\argmax_i P_{ji}(t)=\argmax_{i \in \mathcal{S}} \bm{e}_j^T P(t)=i^*=3$. The figure shows the case for $j=1$, i.e., the CTMC starts from state 1 at time zero.}
    \label{fig:map_osc_tmap}
\end{figure}

The proof of Lemma~\ref{lem:T} is given in Appendix~\ref{appen:lem_T}. In essence, Lemma~\ref{lem:T} states that regardless of the starting state, the MAP estimate of $X(t)$ will be the same beyond a finite time $\tau$. In here, we formally define $\tau^*$ to be the smallest such $\tau$ above which the MAP estimator is constant. We call these CTMCs to have a unique stationary maximum (i.e., $i^*=\argmax_{i\in S}\pi_i$ is unique). Fig.~\ref{fig:map_osc_tmap} illustrates the variation of transition probabilities for one such CTMC.

\begin{definition}
    A CTMC is reversible with respect to measure $\bm{\pi}$ if it satisfies the detailed balance equations $\pi_iq_{ij}=\pi_j q_{ji},$ $\forall i,j\in \mathcal{S}$, and $i\neq j$.
\end{definition}

Throughout the paper, we say that a CTMC is time-reversible, if the CTMC is reversible with respect to its stationary distribution $\bm{\pi}$. Let $\Pi$ be a diagonal matrix with $\Pi_{ii}=\pi_i$. If the CTMC is time-reversible, then $\Pi^{\frac{1}{2}}Q\Pi^{-\frac{1}{2}}$ is symmetric, and hence, can be diagonalized as $UDU^T$, where $D$ is a diagonal matrix whose diagonal elements are the eigenvalues of $Q$ given by $\{-d_1,-d_2,\dots,-d_S\}$. Since $Q\mathbf{1}=\bm{0}$, there is $i_0\in \mathcal{S}$ such that $d_{i_0}=0$. Moreover, we have that $\forall i\neq i_0$, $d_i>0$ \cite{gallager}. Thus, $cI-Q^T$, where $I$ is the identity matrix and $c\in\mathbb{R}$, is invertible $\forall c>0$.  Furthermore, we have that $\bm{u}_i^T\bm{u}_j=0$, $\forall i,j \in \mathcal{S}$ and $i \neq j$, where $\bm{u}_i=\{u_{i1},u_{i2},\dots, u_{iS}\}$ is the $i$th column of the matrix $U$. Hence, for time-reversible CTMCs, we can obtain closed-form expressions for the transition probabilities as stated in Lemma~\ref{lem:Pij}.

\begin{remark}
    For a generic generator matrix $Q$, the real components of the eigenvalues of $Q$ are non-positive \cite{Gershgorin_Circle}. Therefore, for $c\in \mathds{R}^+$, $cI-Q^T$ is an invertible matrix.
\end{remark}

\begin{lemma}\label{lem:Pij}
    For time-reversible CTMCs, the elements of the transition rate matrix are characterized as follows,
    \begin{align}
        P_{ij}(t)=\sqrt{\frac{\pi_j}{\pi_i}}\sum_{k=1}^Su_{ki}u_{kj}e^{-d_kt},
    \end{align}
    where $P_{ij}(t)$ is the probability $X(t)=j$ given $X(0)=i$.
\end{lemma}

The proof of Lemma~\ref{lem:Pij} is given in Appendix~\ \ref{appen:lem:Pij}. Lemma~\ref{lem:Pij} uncovers an important structural characteristic for the MAP estimates of time-reversible CTMCs which we state in  Lemma~\ref{lem:finite_roots}.

\begin{lemma}\label{lem:finite_roots}
    The MAP estimate of a time-reversible CTMC given any starting state, is a piecewise-constant function with finitely many transition points.
\end{lemma}

\begin{figure}[t]
    \centering
    \begin{subfigure}[b]{0.45\textwidth}
        \centering
        \includegraphics[width=0.8\textwidth]{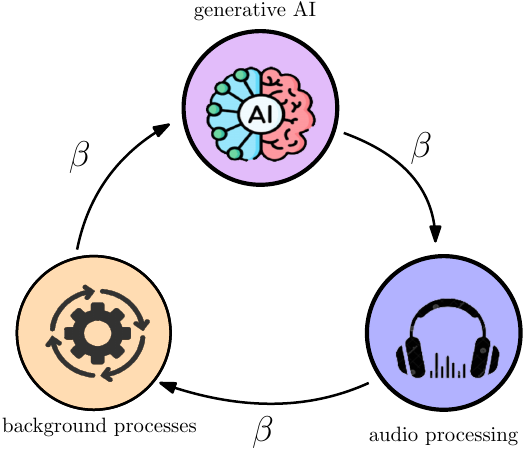}
        \caption{Non time-reversible CTMC.}
        \label{fig:non_tr_ctmc}
    \end{subfigure}
    \hfill 
    \begin{subfigure}[b]{0.45\textwidth}
        \centering
        \includegraphics[width=\textwidth]{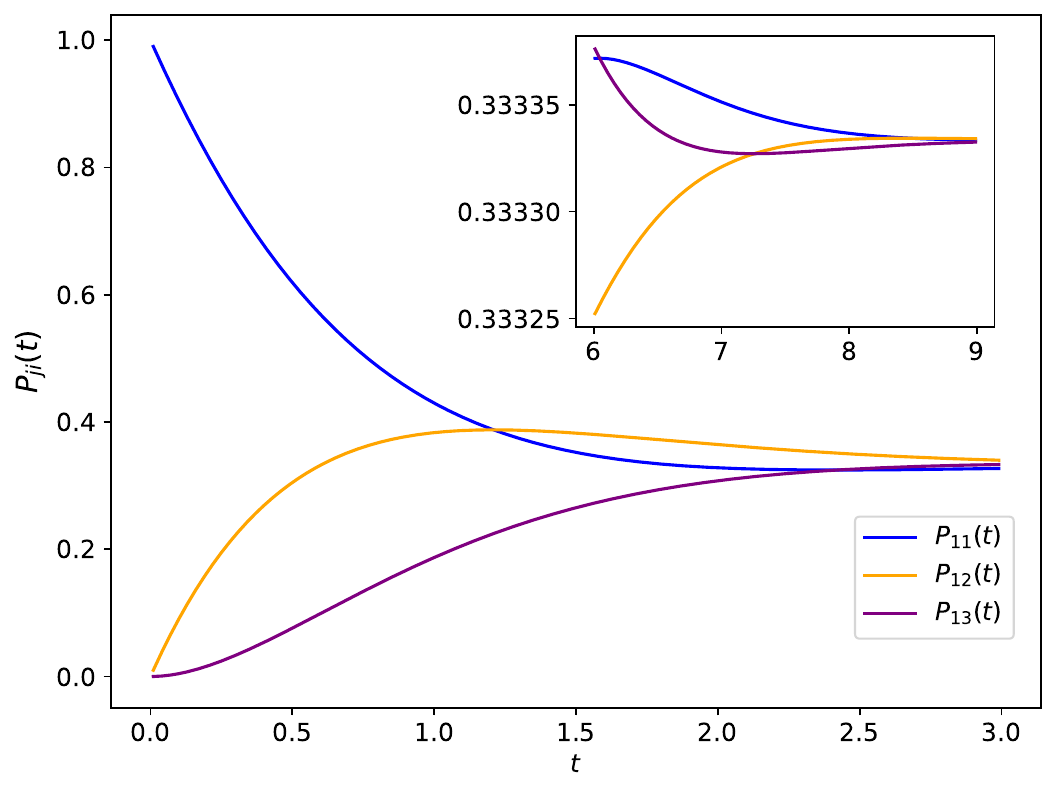}
        \caption{Transition probabilities.}
        \label{fig:osc}
    \end{subfigure}
    \caption{(a) An infinitely oscillating CTMC without a unique stationary maximum. Here, the stationary distribution is $\bm\pi=\{\frac{1}{3},\frac{1}{3},\frac{1}{3}\}$, with no unique $i^*=\argmax_{i\in \mathcal{S}}\pi_i$. (b) There is no time $\tau^*$ after which $\argmax_i P_{ji}(t)$ is constant, in fact, it is infinitely oscillating.}
    \label{fig:map_osc}
\end{figure}

The proof of Lemma~\ref{lem:finite_roots} is given in Appendix~\ref{appen:lem:finite_roots}. This characteristic is not only limited to time-reversible CTMCs. It can indeed be generalized to any CTMC whose generator matrix is similar to a diagonal matrix of real eigenvalues. For a generic CTMC, this may not be always true. Fig.~\ref{fig:map_osc} illustrates one such scenario where the depicted CTMC is not time-reversible and has complex eigenvalues. As a result, the transition probabilities tend to infinitely oscillate, and hence, the MAP estimator is infinitely oscillating. These oscillations are solely driven by the complex eigenvalues of the generator matrix. However, regardless whether the CTMC is non time-reversible or whether its generator matrix has complex eigenvalues, if it has a unique stationary maximum as in Fig.~\ref{fig:map_osc_tmap}, we have that the MAP estimator has finitely many transition points as a consequence of Lemma~\ref{lem:T}.

Next, let $R(t)$ denote an absorbing CTMC with states $\{1,2,\dots,\Gamma,\Gamma+1\}$ whose state transitions involve only transition from the $i$th state to the $(i+1)$th state with rate $\lambda$ where state $\Gamma+1$ is the absorbing state. We call this structure  an \emph{Erlang chain}. Let $\tau^{(\lambda)}$ denote the sum of $\Gamma$ independent exponentially distributed random variables with rate $\lambda$. Then, $\tau^{(\lambda)}$ follows an Erlang distribution with shape parameter $\Gamma$ and rate $\lambda$. Moreover, $\tau^{(\lambda)}$ models the time to reach the absorbing state starting from state 1. Additionally, we have that $\e[\tau^{(\lambda)}]=\frac{\Gamma}{\lambda}$ and $\text{Var}[\tau^{(\lambda)}]=\frac{\Gamma}{\lambda^2}$. Now, consider the random variable $\tau^{(\lambda \Gamma)}$. We have $\e[\tau^{(\lambda\Gamma)}]=\frac{1}{\lambda}$ and $\text{Var}[\tau^{(\lambda \Gamma)}]=\frac{1}{\lambda^2\Gamma}$. Therefore, $\tau^{(\lambda\Gamma)}$ converges to the deterministic value $\frac{1}{\lambda}$ in the mean square sense, and hence, in distribution as $\Gamma\to\infty$. 

\section{System Model} \label{sec:sys_model}
Let $X(t)$ be a finite-state irreducible CTMC with transition rate matrix $Q$. This CTMC is monitored through a query-based sampling procedure, where the remote monitor sends queries to the CTMC, requesting its status, at exponentially distributed intervals with a fixed rate of $\mu$. Upon receiving a query, the CTMC instantaneously transmits its current state to the remote monitor, which then updates its estimate $\hat{X}(t)$ of the CTMC. We assume that the initial states $X(0)$ and $\hat{X}(0)$ are distributed according to the stationary distribution $\bm\pi$ of the CTMC. Traditionally, the estimate $\hat{X}(t)$ is kept constant during inter-query intervals and we call such estimators single-stage estimators since the estimator is kept fixed between two successive queries. The most natural single-stage estimator is the martingale estimator (ME), where the estimate is fixed to the most recently received update until the next query. 

\begin{figure}[t]
    \centering
    \includegraphics[width=\textwidth]{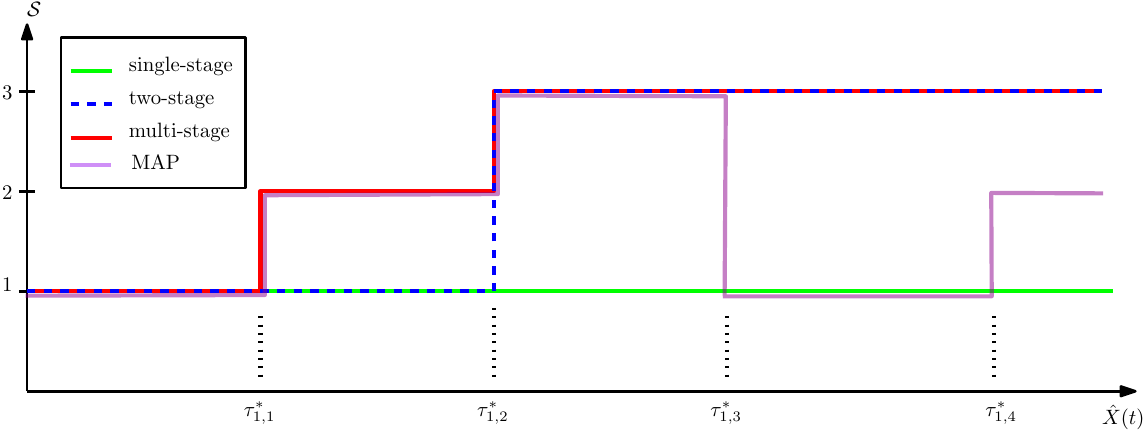}
    \caption{Evolution of the estimate $\hat{X}(t)$ at the remote monitor before receiving an update given $\hat{X}(0)=1$. Here, $\tau^*_{1,k}$ denotes the $k$th transition point of the MAP estimator.}
    \label{fig:multi_stage}
\end{figure}

We transcend past these single-stage estimators and enable our estimators to evolve during the inter-query periods. We call such estimators \emph{structured estimators}, where as the name suggests, the estimator is allowed to change between two queries based on a predetermined structure without having to receive a new status update. Among these structured estimators, the simplest structure that can be followed is the one where the estimator changes only once during the inter-query period. We call such estimators \emph{two-stage structured estimators}. Within these two-stage structured estimators, the most common configuration is the one where the estimator shifts from the martingale estimate to the MAP estimate at one time point. If the estimator is allowed to change more than once between two queries, we call them \emph{multi-stage structred estimators}. Ideally, we would like these changes or shifts to be aligned with the MAP estimate of the system. Fig.~\ref{fig:multi_stage} illustrates the evolution of these structured estimators for the non-time-reversible CTMC with its infinitely oscillating MAP estimator depicted in Fig.~\ref{fig:map_osc}(a). In general, multi-stage estimators will try to closely align with the fluctuations of the MAP estimator as we increase the number of stages. In this work, we will be analyzing the MBF when such structured estimators are employed.

Now, we will introduce the various estimators studied in this work. First, we will formally define the most commonly used single-stage estimator which is the ME. Then, we will introduce several simple two-stage estimators. Finally, we will rigorously define our multi-stage estimator, that we consider in this work.

\subsection{Martingale Estimator}
Let $\hat{X}_M(t)$ denote the ME at the monitor. Here, $\hat{X}_M(t)$ will retain the previously received state as its estimate until updated by a new sample. Let $G(t)$ denote the last time the monitor received an update. Then, we define $\hat{X}_M(t)$ as follows,
\begin{align}
    \hat{X}_M(t)=X(G(t)).
\end{align}

\subsection{Simple Two-Stage Structured Estimators} \label{sec:TSSE}
Now, we formally describe several simple two-stage structured estimators. A simple two-stage structured estimator is an estimator which shifts between the ME and the state $i^*$ based on some structure. In particular, we will either have a deterministic or random threshold, where the estimator will shift from the ME to the value $i^*$, upon exceeding this threshold. Here, we call the above definition of the two-stage estimator \emph{simple}, since the threshold and the state to which our estimator transitions after exceeding the threshold, are both independent of the most recently observed state of the CTMC. Only the value of the estimator in the first stage depends on the observed state. Despite their simplicity, these simple two-stage estimators are analytically rich and are well-defined for CTMCs with a unique stationary maximum, where they provide an unadorned two-stage approximation for the MAP estimator. Even for CTMCs without a unique stationary maximum, they can still be used, with a slight abuse of notation and at the expense of freshness, by redefining $i^*=\max\{\argmax_{i\in \mathcal{S}}\pi_i\}$ in case there are multiple maximal elements in $\bm\pi$. The formal definition of a generic two-stage estimator, where both the threshold and the value to which the estimator transitions after exceeding the threshold, are dependent on the most recently observed state of the CTMC, will be provided when we define our multi-stage estimators in Section~\ref{sec:mult_se}. Here, we continue by defining three simple two-stage estimators.

\subsubsection{Exponential Estimator (EXPE)}\label{sec:exp_est}
Denote by $\hat{X}_{e,\lambda}(t)$ the exponential estimator. In here, once the monitor receives a new sample, it will start an exponential clock whose rate is $\lambda$. The monitor will retain the received sample as its estimate until a new sample is obtained or until the exponential timer runs out. If the exponential timer runs out, it will change its estimate to $i^*$,  which is the MAP estimate of $X(t)$ if we did not receive an update for a time period beyond $\tau^*$. It will then retain this $i^*$ state until a new sample is obtained. Each time a new sample is obtained, the exponential timer is reset. Let $S_i$ denote the $i$th sampling time. Since we sample at exponential intervals, we have $(S_{i+1}-S_i)\sim Exp(\mu)$. Let $\tilde{S}_i\sim Exp(\lambda)$ denote the realization of the exponential clock for the $i$th sample. Then, $\hat{X}_{e,\lambda}(t)$ can be defined as follows,
\begin{align}
    \hat{X}_{e,\lambda}(t)=\begin{cases}
        X(G(t)),& \text{if}~ t\in (S_i,\min\{S_i+\tilde{S}_i,S_{i+1}\}],\\
        i^*, & \text{if}~t\in (\min\{S_i+\tilde{S}_i,S_{i+1}\},S_{i+1}).
    \end{cases}
\end{align}

\subsubsection{Erlang Estimator (ERLE)}
\begin{figure}[t]
    \centering
    \includegraphics[width=0.8\textwidth]{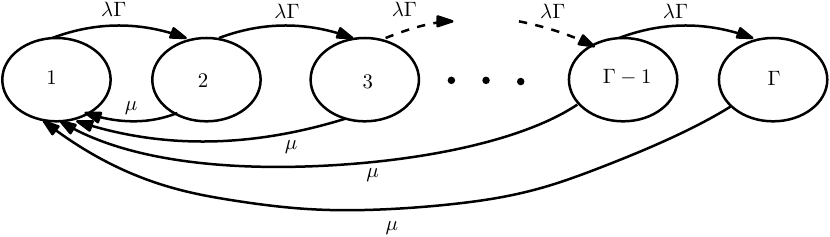}
    \caption{State transition diagram for $Y(t)$.}
    \label{fig:erlang_chain}
\end{figure}

Let $Y(t)$ be a CTMC independent of $X(t)$, with $\Gamma$ states where the state transitions occur only from the $k$th state to the $(k+1)$th state with rate $\lambda \Gamma$ and from the $k$th state to state $1$ with rate $\mu$. State $\Gamma$ can only transition to state $1$ with rate $\mu$; see Fig.~\ref{fig:erlang_chain}. Starting with $Y(0)=1$, the time to reach state $\Gamma$ when  $\mu=0$ is governed by an Erlang distribution whose rate and shape parameters are $\lambda \Gamma$ and $\Gamma-1$. Denote by $\hat{X}_{\Gamma,\lambda}(t)$, the ERLE. Then, $\hat{X}_{\Gamma,\lambda}(t)$ is defined as follows,
\begin{align}
    \hat{X}_{\Gamma,\lambda}(t)=\begin{cases}
        X(G(t)),& \text{if}~Y(t)\neq \Gamma,\\
        i^*,&\text{if}~Y(t)=\Gamma.
    \end{cases}
\end{align}

\subsubsection{$\tau$-MAP Estimator}
Let $\hat{X}_\tau(t)$ denote the $\tau$-MAP estimator and let $\delta(t)$ be the time elapsed since the last sample was taken, i.e., $\delta(t)=t-G(t)$, which we will also refer to as the age of the estimator. In here, the monitor retains the previous sample until a new sample is obtained or until $\delta(t)>\tau$, i.e., the age of the estimator exceeds threshold $\tau$. If $\delta(t)>\tau$, then it will change its estimate to $i^*$. Thus, $\hat{X}_\tau(t)$ is defined as follows,
\begin{align}
    \hat{X}_\tau(t)=\begin{cases}
        X(G(t)),& \text{if}~ \delta(t)\leq \tau,\\
        i^*,& \text{if}~ \delta(t)> \tau.
    \end{cases}
\end{align}

\begin{remark}\label{rem:convergence}
    As $\Gamma \to \infty$, $\hat{X}_{\Gamma,\frac{1}{\tau}}(t)$ converges to  $\hat{X}_\tau(t)$ in distribution.
\end{remark}

\begin{remark}
    The interest in the study of two-stage estimators, such as EXPE and ERLE, where the shifts in the estimator occur at random time periods, is mainly due to their simplicity when analyzing them. In fact, we show that, having deterministic shifts, such as in the $\tau$-MAP estimator, can yield superior MBF gains.
\end{remark}

\subsection{Multi-Stage Structured Estimators} \label{sec:mult_se}
In here, we consider an estimator that approximates the MAP estimator as a piecewise-constant function with finitely many transition stages, where in each stage, the estimator assumes the state which maximizes the MBF metric. As opposed to the simple two-stage estimators where we used a single universal threshold independent of the observed state, in here, we will enforce multiple state-dependent deterministic thresholds for each observed state, such that the estimator will remain constant between any two thresholds. Now, we formally describe the multi-stage estimator considered in this work.

\subsubsection{$p$-MAP Estimator}
This is an extension of the $\tau$-MAP estimator, where we consider multiple intermediate transition stages for the estimator. Suppose our most recent sample indicates that the CTMC was in state $i$. Then, we will allow our estimator to evolve through at most $K_i$ stages before the next sampling instance, based on the age of the estimator. In here, a stage is an interval of time where our estimator remains constant. Let us  represent these intervals using the sequence of non-negative real numbers  $\{\tau_{i,0},\tau_{i,1},\dots,\tau_{i,K_i}\}$   with $\tau_{i,0}=0$, $\tau_{i,K_i}=\infty$ and $\tau_{i,k}\leq \tau_{i,k+1}$, where the interval $[\tau_{i,k},\tau_{i,k+1})$ represents the $k$th stage of our estimator if the most recently sampled state was $i$. As we wait for the next sample, the age of our estimator, $\delta(t)$, increases. If $\delta(t)$ belongs to the interval $[\tau_{i,k},\tau_{i,k+1})$, i.e., the $k$th stage, then the estimator will assume the value $\Gamma_{i,k}$. Therefore, as the age of our estimator increases, the estimator will assume different values and evolve through multiple stages. The number of stages, the duration of the stages, and the value the estimator assumes in each stage, depend on the most recent sample, i.e., the most recent observation. Now, the $p$-MAP estimator, denoted by $\hat{X}_P(t)$, can be formally defined as follows,
\begin{align}
    \hat{X}_p(t)=\sum_{k=1}^{K_{i(t)}}\Gamma_{i(t),k}\mathds{1}\left\{\delta(t)\in \left[\tau_{i(t),k-1},\tau_{i(t),k}\right]\right\},
\end{align}
where  $\Gamma_{i,k}=\argmax_j\int_{\tau_{i,k-1}}^{\tau_{i,k}}P_{ij}(t)e^{-\mu t}\,\dd{t}$ for $k>1$ with $\Gamma_{i,1}=i$  and $i(t)=X(G(t))$. Sometimes we will use $i_k$ instead of $\Gamma_{i,k}$ for brevity. In here, $\Gamma_{i,k}$ is explicitly chosen to maximize the MBF metric which will be evident in Section~\ref{sec:Fresh_MSEs}. We resort to this definition for the $p$-MAP estimator to facilitate the approximation of the MAP estimate with the desired number of intermediate stages. Moreover, this definition enables us to go beyond time reversibility and generalize to any CTMC (even the ones with infinitely oscillating MAP estimators). However, in this work, we are mainly focused on time-reversible CTMCs. In fact, for a time-reversible CTMC, if we replace $\tau_{i,k}$ with $\tau_{i,k}^*$ which is the $k$th transition point of the MAP estimate with $X(0)=i$, then $i_k$s will be independent of $\mu$ and the $p$-MAP estimator will be equivalent to the MAP estimator. If $K_i=2$ for all $i$, we get the formal definition of a generic two-stage estimator. In fact, when  $K_i=2$ and $i_2=i^*$ for all $i$ with $\tau_{i,1}=\tau$, then the definitions of the $p$-MAP estimator and the $\tau$-MAP estimator coincide for CTMCs with a unique stationary maximum. 

\section{Freshness Under State-Independent Sampling} \label{sec:main_results}
In this section, we present our main analytical results for the MBF metric for the considered structured estimators, when employing a fixed sampling rate $\mu$ to sample the CTMC, independent of the observed state. We will extend our results to the case of last observed state-dependent sampling policies in Section~\ref{sec:state_dep}.

\subsection{Freshness of Simple Two-Stage Estimators}
Let $\text{tr}(\cdot)$ be the trace operator and denote by $\e[\Delta_{M}]$, $\e[\Delta_{e,\lambda}]$, $\e[\Delta_{\Gamma,\lambda}]$ and $\e[\Delta_{\tau}]$ the MBF obtained under the martingale (ME), exponential (EXPE), Erlang (ERLE) and $\tau$-MAP estimators, respectively. In here, we will only provide analytical results for the latter three estimators. The analysis for the ME can be found in \cite{nail_QS}. 

\begin{theorem}\label{thrm:exp_BF}
    Under EXPE, $\e[\Delta_{e,\lambda}]$ is given by,
    \begin{align}
       \e[\Delta_{e,\lambda}]= \mu\text{tr}\left(\tilde{Q}_0\Pi\right)+\frac{\lambda}{\mu+\lambda}\pi_{i^*},\label{eqn:exp_BF}
    \end{align}
    where $\tilde{Q}_0=\left((\mu+\lambda)I-Q^T\right)^{-1}$.
\end{theorem}

The proof of Theorem~\ref{thrm:exp_BF} follows arguments similar to those of the ERLE, and hence its proof is omitted, and ERLE proof (proof of Theorem~\ref{thrm:erlang_BF}) is given in detail.

\begin{remark}
  $\e[\Delta_{e,0}]$ simplifies to the exact expression for the MBF metric under a ME obtained in \cite{nail_QS}. 
\end{remark}

\begin{theorem}\label{thrm:erlang_BF}
    Under ERLE, $\e[\Delta_{\Gamma,\lambda}]$ is given by,
    \begin{align}
       \e[\Delta_{\Gamma,\lambda}]= \mu\sum_{k=1}^{\Gamma-1}(\lambda \Gamma)^{k-1}\text{tr}\left(\tilde{Q}^{k}\Pi\right)+\frac{(\lambda \Gamma)^{\Gamma-1}}{(\mu+\lambda \Gamma)^{\Gamma-1}}\pi_{i^*},\label{eqn:erlang_BF}
    \end{align}
    where $\tilde{Q}=\left((\mu+\lambda \Gamma)I-Q^T\right)^{-1}$.
\end{theorem}

The proof of Theorem~\ref{thrm:erlang_BF} is given in Appendix~\ref{appen:erlang_BF}. Next, we analyze the expression in Theorem~\ref{thrm:erlang_BF} explicitly for time-reversible CTMCs.

\begin{corollary}\label{cor:erl_tr}
    If $X(t)$ is time-reversible, then under ERLE, $\e[\Delta_{\Gamma,\lambda}]$ is given by,
    \begin{align}
        \e[\Delta_{\Gamma,\lambda}]=\sum_{i=1}^S\frac{a_i\mu}{d_i+\mu}\left( 1-\left(\frac{\lambda\Gamma}{d_i+\mu+\lambda\Gamma}\right)^{\Gamma-1}\right)+\frac{(\lambda \Gamma)^{\Gamma-1}}{(\mu+\lambda \Gamma)^{\Gamma-1}}\pi_{i^*},\label{eqn:TR_BF_K}
    \end{align}
    where $-d_i$ are the eigenvalues of $Q$ and $a_i=\sum_{j=1}^S\pi_ju_{ij}^2$.
\end{corollary}

The proof of Corollary~\ref{cor:erl_tr} is given in Appendix~\ref{appen:cor_erl_tr}. Now, from \eqref{eqn:exp_BF} and \eqref{eqn:erlang_BF}, note that $\e[\Delta_{e,\lambda}]=\e[\Delta_{2,\frac{\lambda}{2}}]$. Therefore, by setting $\Gamma=2$ and $\lambda=\frac{\lambda}{2}$ in \eqref{eqn:TR_BF_K}, we can obtained the corresponding counterpart of Corollary~\ref{cor:erl_tr} for EXPEs.

\begin{corollary}
    If $X(t)$ is time-reversible, then under EXPE, $\e[\Delta_{e,\lambda}]$ is given by,
    \begin{align}
        \e[\Delta_{e,\lambda}]=\sum_{i=1}^S\frac{a_i\mu}{d_i+\mu+\lambda}+\frac{\lambda}{\mu+\lambda}\pi_{i^*}.
    \end{align}
\end{corollary}

By Remark~\ref{rem:convergence} and dominated convergence \cite{koralov_sinai}, we have that $\e[\Delta_\tau]=\e[\lim_{\Gamma\to\infty}\Delta_{\Gamma,\frac{1}{\tau}}]=\lim_{\Gamma \to \infty}\e[\Delta_{\Gamma,\frac{1}{\tau}}]$. This is valid for any generic CTMC with a unique stationary maximum. 

Next, from Corollary~\ref{cor:erl_tr} and using the fact that $\lim_{n\to\infty}\left(1+\frac{1}{n}\right)^n=e$, we obtain the expression for $\e[\Delta_\tau]$ for time-reversible CTMCs.

\begin{corollary}
    If $X(t)$ is time-reversible, then under the $\tau$-MAP estimator, $\e[\Delta_{\tau}]$ is given by,
    \begin{align}
    \e[\Delta_{\tau}]=\sum_{i=1}^S\frac{a_i\mu}{d_i+\mu}\left(1-e^{-(d_i+\mu)\tau}\right)+e^{-\mu \tau}\pi_{i^*}.
    \end{align}
\end{corollary}

Next, we present an important relationship between the martingale and $\tau$-MAP estimators. The proof of Theorem~\ref{thrm:mart_v_map} is given in Appendix~\ref{appen:mart_v_map}.

\begin{theorem}\label{thrm:mart_v_map}
   Let $\tau^*$ be as defined in Lemma~\ref{lem:T} and $\argmax_{i\in \mathcal{S}}\pi_i$ be unique. Then, the following relation holds,
   \begin{align}
        \e[\Delta_M] \leq \e[\Delta_{\tau^*}].
   \end{align}
\end{theorem}

\subsection{Freshness of Multi-Stage estimators} \label{sec:Fresh_MSEs}
Now, we present our main analytical results for the $p$-MAP estimator. The techniques developed in this section can readily be used to analyze the MBF metric of any estimator, including the ones described in Section~\ref{sec:TSSE}. In fact, we can directly translate our results in this section for the $\tau$-MAP estimator and the ME. 

Now, to state our results, in Theorem~\ref{thrm:fresh}, we first show that MBF under any estimator can be expressed using the average time the estimator was fresh between two sampling instances.

\begin{theorem}\label{thrm:fresh}
    The MBF, $\e[\Delta]$, is given by,
    \begin{align}
        \e[\Delta]=\mu\sum_{i=1}^S\pi_i\e[F_i],
    \end{align}
    where $\e[F_i]$ is the expected portion of time the estimator is fresh starting from state $i$ until the next sample is taken.
\end{theorem}

The proof of Theorem~\ref{thrm:fresh} is given in Appendix~\ref{appen:thrm:fresh}. Next, we explicitly find the expression for $\e[F_i]$ for any given estimator in Theorem~\ref{thrm:F_i}. The proof of Theorem~\ref{thrm:F_i} is given in Appendix~\ref{appen:thrm:F_i}.

\begin{theorem}\label{thrm:F_i}
    The expected portion of time  the estimator was fresh starting from state $i$, is given by,
    \begin{align}
        \e[F_i]=\int_{0}^\infty P_{i\hat{X}(t)}(t)e^{-\mu t}\,\dd{t}.\label{eqn:Fint}
    \end{align}
\end{theorem}

\begin{corollary}\label{cor:pmap}
    For a time-reversible CTMC, the MBF metric under the $p$-MAP estimator is given by,
    \begin{align}
    \e[\Delta_P]=\sum_{i=1}^S\sum_{j=1}^S\sum_{k=1}^{K_i}a_{i,j,k}\frac{\mu}{\mu+d_j}\left(e^{-(d_j+\mu)\tau_{i,k-1}}-e^{-(d_j+\mu)\tau_{i,k}}\right), \label{eqn:pmap}
    \end{align}
    where $a_{i,j,k}=\sqrt{\pi_i\pi_{i_k}}u_{ji}u_{ji_k}$.
\end{corollary}

The proof of Corollary~\ref{cor:pmap} directly follows by evaluating the integral in \eqref{eqn:Fint}. The MBF of the $\tau$-MAP estimator  directly follows by setting $K_i=2$, $i_2=i^*$ and  $\tau_{i,1}=\tau$ in \eqref{eqn:pmap}, and the MBF for the ME can be obtained by setting $\tau=\infty$.

\section{Freshness Under State-Dependent Sampling} \label{sec:state_dep}
In this section, we analyze the MBF when a state-dependent sampling rate policy is employed instead of sampling at a fixed rate of $\mu$ in all states. In particular, we will consider that the remote monitor will sample the CTMC at rate of $\mu_i>0$, if $\hat{X}_M(t)=i$, i.e., if the last sample value was $i$. We will first find the MBF of an ME and will later use its result to extend it to general estimators, such as the $p$-MAP estimator. To distinguish between  state-dependent sampling and fixed sampling, we use $\tilde{\Delta}$ to denote the MBF under a state-dependent policy.

\subsection{Martingale Estimator}
Let $\tilde{Z}(t)=(X(t),\hat{X}_M(t))$. Then, $\tilde{Z}(t)$ is a finite-state irreducible two dimensional CTMC. Hence, its stationary distribution denoted by the column vector $\bm\psi=\{\psi_{ij}\}_{i,j\in \mathcal{S}}$, exists. Let $Q_M$ be its generator matrix whose elements are denoted by $Q_M[s_1,s_2]$, where $s_1,s_2\in \mathcal{S}\cross \mathcal{S}$. Then, $Q_M$ can be fully characterized as follows,
\begin{align}
    Q_M[(i,j),(k,l)]=\begin{cases}
        \mu_j,&\text{if}~k=l=i\neq j,\\
        q_{ik},&\text{if}~k\neq i ~\text{and}~l=j,\\
        -q_i+\mu_j, &\text{if}~k=i\neq j=l,\\
        -q_i,&\text{if}~k=l=i=j,\\
        0,& \text{otherwise}.
    \end{cases}
\end{align}
Then, $\bm \psi$ is the unique solution that satisfies the global balanced equations $\bm\psi^TQ_M=\bm{0}$ and $\bm\psi^T\mathbf{1}=1$. Let $\tilde{Q}_M$ be the transpose of matrix $Q_M$ with the last row replaced by a row of ones and let $\bm{e}_{S^2}$ be a column vector of all zeros except at the last index. Then, $\bm\psi$ can be found as follows,
\begin{align}
    \bm\psi= \tilde{Q}_M^{-1} \bm{e}_{S^2}.
\end{align}
Then, the MBF under an ME can be found as,
\begin{align}
    \e[\tilde{\Delta}_M]=\sum_{i \in \mathcal{S}}\psi_{ii}.
\end{align}

\subsubsection{Binary CTMC}
\begin{figure}[t]
    \centering
    \includegraphics[width=0.4\textwidth]{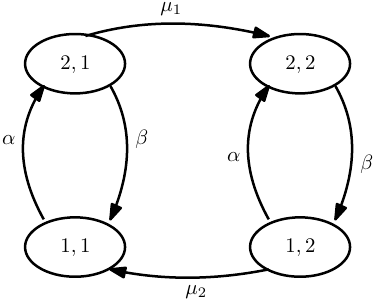}
    \caption{$\tilde{Z}(t)$ for a binary CTMC.}
    \label{fig:binary_ctmc}
\end{figure}

As an example, we explicitly give the expression for the resulting MBF under an ME for a binary CTMC, to highlight some of the analytical complexities that may arise in the later sections. Let the binary CTMC transition between states $1$ and $2$ with rates $\alpha$ and $\beta$ as illustrated in Fig.~\ref{fig:binary_ctmc}. Then, solving $\bm\psi$ as described above yields,
\begin{align}
    \psi_{11}&=\frac{\beta\mu_2(\beta+\mu_1)}{(\alpha+\beta)(\mu_1\mu_2+\alpha\mu_1+\beta\mu_2)},\\
    \psi_{21}&=\frac{\alpha\beta\mu_2}{(\alpha+\beta)(\mu_1\mu_2+\alpha\mu_1+\beta\mu_2)},\\
    \psi_{12}&=\frac{\alpha\beta\mu_1}{(\alpha+\beta)(\mu_1\mu_2+\alpha\mu_1+\beta\mu_2)},\\
    \psi_{22}&=\frac{\alpha\mu_1(\mu_2+\alpha)}{(\alpha+\beta)(\mu_1\mu_2+\alpha\mu_1+\beta\mu_2)},
\end{align}
resulting in the MBF,
\begin{align}
    \e[\tilde{\Delta}_M]&=\frac{(\alpha+\beta)\mu_1\mu_2+\beta^2\mu_2+\alpha^2\mu_1}{(\alpha+\beta)(\mu_1\mu_2+\alpha\mu_1+\beta\mu_2)}. \label{eqn:mart_fresh_exp}
\end{align}
Note, how $\e[\tilde{\Delta}_M]$ and $\psi_{ij}$s are not necessarily jointly convex functions with respect to $\mu_1$ and $\mu_2$. Hence, finding the optimal values for $\mu_1$ and $\mu_2$, when subjected to a sampling rate constraint, can be challenging using conventional approaches.

\subsection{General Estimators}
Now, we will find the MBF of the $p$-MAP estimator. We will first show that the MBF of a generic (i.e., general) estimator can be computed using the average time the estimator was fresh between sampling instances, $\e[\tilde{F}_{i,\mu_i}]$ and $\bm\psi$. Here, $\e[\tilde{F}_{i,\mu_i}]$ is obtained by replacing $\mu$ with $\mu_i$ in \eqref{eqn:Fint}.

\begin{theorem}\label{thrm:fresh_mui}
    The MBF $\e[\tilde{\Delta}]$ under a state-dependent sampling policy is given by,
    \begin{align}
        \e[\tilde{\Delta}]=\sum_{i=1}^S\mu_i\tilde{\pi}_i\e[\tilde{F}_{i,\mu_i}],
    \end{align}
    where $\tilde{\pi}_i=\sum_{j\in \mathcal{S}}\psi_{ji}$ is the proportion of time the ME was in state $i$.
\end{theorem}

\begin{corollary}\label{cor:pmap_mui}
    For a time-reversible CTMC, the MBF for the $p$-MAP estimator under a state-dependent sampling policy is given by,
    \begin{align}
    \e[\tilde{\Delta}_P]=\sum_{i=1}^S\sum_{j=1}^S\sum_{k=1}^{K_i}\tilde{a}_{i,j,k}\frac{\mu_i}{\mu_i+d_j}\left(e^{-(d_j+\mu_i)\tau_{i,k-1}}-e^{-(d_j+\mu_i)\tau_{i,k}}\right), \label{eqn:pmap_mui}
    \end{align}
    where $\tilde{a}_{i,j,k}=\sqrt{\frac{\pi_{i_k}}{\pi_i}}\tilde{\pi}_iu_{ji}u_{ji_k}$.
\end{corollary}

\begin{corollary}\label{cor:tmap_mui}
    The MBF under the $\tau$-MAP estimator is given by,
    \begin{align}
    \e[\tilde{\Delta}_\tau]=\sum_{i=1}^S\sum_{j=1}^S\frac{\mu_i}{d_j+\mu_i}\left(\tilde{\pi}_iu_{ji}^2-\tilde{b}_{i,j}e^{-(d_j+\mu_i)\tau}\right),\label{eqn:tmap_mui}
    \end{align}
    where $\tilde{b}_{i,j}=\tilde{\pi}_iu_{ji}^2-\sqrt{\frac{\pi_{i^*}}{\pi_i}}\tilde{\pi}_iu_{ji}u_{ji^*}$.
\end{corollary}

The proof of Theorem~\ref{thrm:fresh_mui} is given in Appendix~\ref{appen:thrm:fresh_mui}. The proofs of Corollary~\ref{cor:pmap_mui} and Corollary~\ref{cor:tmap_mui} follow by evaluating the integral in \eqref{eqn:Fint} with $\mu$ replaced by the respective $\mu_i$. Moreover, following the proof of Theorem~\ref{thrm:fresh_mui}, we can further show that the the average sampling rate under this policy denoted by $\omega$, is given by $\omega=\sum_{i\in \mathcal{S}}\tilde{\pi}_i\mu_i$. Note that, the sampling rate, regardless of the estimator, is always is equal to that of the ME. 

\subsection{Optimal State-Dependent Sampling Policy}
When employing a state-dependent sampling policy, a natural question to ask is how one can optimally allocate the sampling rates while being subjected to an overall average sampling budget. This leads us to the following optimization problem,
\begin{maxi}
    {\mu_i}{\e[\tilde{\Delta}]}
    {\label{eqn:opt_state_dep}}
    {}
    \addConstraint{\sum_{i=1}^S \mu_i\tilde{\pi}_i}{\leq \Omega},
\end{maxi}
where $\Omega$ is the maximum average sampling rate allowed. In here, we will solely focus on time-reversible CTMCs. Both the objective and the constraint of the above optimization problem are non-convex functions of $\mu_i$s. Therefore, finding closed-form solutions and analytically optimizing it, is a very difficult task. To circumnavigate these challenges, we formulate the problem as an SMDP and will use a policy iteration algorithm to find the optimal sampling rate for each state.

\begin{remark}
    The results in this section are not limited to time-reversible CTMCs or $p$-MAP estimators. In fact, the SMDP framework used in this section is readily applicable  to any CTMC with any generic estimator as long as the integral in $\e[\tilde{F}_{i,\mu_i}]$ can be computed efficiently.
\end{remark}

\subsubsection{SMDP Formulation}
Now, we construct the SMDP for solving the optimization problem in \eqref{eqn:opt_state_dep}. Instead of directly solving \eqref{eqn:opt_state_dep}, we will first convert it to an unconstrained problem with the use of a Langragian multiplier. Further, to reduce the size of the action space of the SMDP, we will restrict the feasible $\mu_i$s to be bounded from above and below. Moreover, for the SMDP, we enable the use of randomized policies by slightly deviating from the original problem which was restricted to the class of deterministic policies. This enables us to tackle a larger class of rate allocation policies generalizing our problem. 

\begin{figure}[t]
    \centering
    \includegraphics[width=0.5\textwidth]{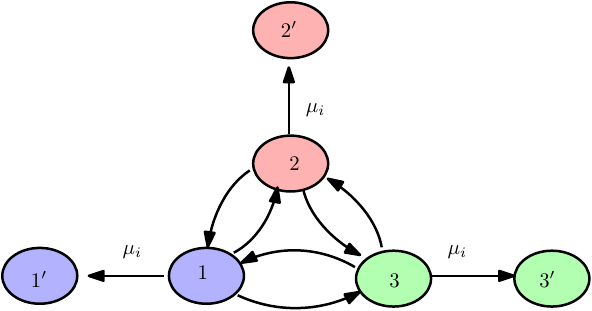}
    \caption{Absorbing Markov chain $Z_i(t)$ for a CTMC with 3 states.}
    \label{fig:amc}
\end{figure}

Now, our optimization problem reduces to the following,
\begin{maxi}
    {\rho_l<\mu_i<\rho_u}{\e[\tilde{\Delta}]-\gamma\sum_{i=1}^S \mu_i\tilde{\pi},}
    {\label{eqn:opt_uncons_lag}}
    {}
\end{maxi}
where $\rho_l<\Omega$ and is chosen very close to zero and $\rho_u$ is chosen well above $\Omega$. The SMDP can be fully characterized using the tuple $(\mathcal{S,A,P,R,H)}$ defined below.
\begin{itemize}
    \item The \emph{state space} $\mathcal{S}=\{1,2,\dots,S\}$ is the state of our ME, $\hat{X}_M(t)$.
    \item The \emph{action space} $\mathcal{A}=[\rho_l,\rho_u]$ is the set of feasible sampling rates.
    \item The \emph{transition function} $\mathcal{P}:\mathcal{S\cross A\cross S}\to [0,1]$ defines the transition probabilities between states based on the selected action. In particular, $\mathcal{P}(s,a,s')$ denotes the probability of transitioning to state $s'$ if the action $a$ was selected at state $s$.
    \item The \emph{reward function} $\mathcal{R}:\mathcal{S\cross A}\to \mathds{R}$ defines the average reward obtained in a  state based on the action selected. Here, $\mathcal{R}(s,a)$ denotes the average reward obtained by selecting action $a$ in state $s$. When in state $i$, if we choose $\mu_i$ as the sampling rate, then $\mathcal{R}(i,\mu_i)=\e[\tilde{F}_{i,\mu_i}]-\gamma$.
    \item The \emph{sojourn times} $\mathcal{H}$ defines the average time the process stays in a particular state based on the action selected. In particular, $\mathcal{H}(s,a)$ denotes the average time the process stays in state $s$ if the action $a$ was selected. For example, if we choose rate $\mu_i$ when in state $i$, then $\mathcal{H}(i,\mu_i)=\frac{1}{\mu_i}$, since the ME changes only when a new sample is obtained.
\end{itemize}
Now, to find the transition probabilities $\mathcal{P}(s,a,s')$, we will construct an absorbing Markov chain and find its absorption probabilities. For example, suppose our SMDP is in state $i$ (i.e., $\hat{X}_M(t)=i$) and we want to find the transition probabilities when action $\mu_i$ is chosen. For this, we will construct the  CTMC $Z_i(t)$, whose states are $\{1,2,\dots ,S,1',2',\dots,S'\}$, where $1'$ to $S'$ denote absorbing states (see Fig.~\ref{fig:amc}). Note that, once a new sample is obtained, the SMDP transitions to a new state and until a new sample is obtained, it will be sampling at the same sampling rate. Hence, the generator matrix $Q_i$ of $Z_i(t)$ is given by,
\begin{align}
    Q_i=\begin{bmatrix}
        Q-\mu_iI_S&\mu_iI_S\\
        0_{S\times S}&0_{S\times S}
    \end{bmatrix},
\end{align}
where $I_S$ is an identity matrix of size $S$, and $0_{S\times S}$ is a matrix of all zeros of size $S\times S$. Now, the transition probabilities can be obtained by the absorbing probabilities of states $1'$ to $S'$, starting from state $i$. In particular, $\mathcal{P}(i,\mu_i,j)$ is equal to the absorption probability of state $j'$. Let $\mathcal{P}(i,\mu_i)=\{\mathcal{P}(i,\mu_i,1),\mathcal{P}(i,\mu_i,2),\dots,\mathcal{P}(i,\mu_i,S)\}$ denote the vector of absorption probabilities. Then, from \cite[Lemma~3]{revMax}, we have  that,
\begin{align}
    \mathcal{P}(i,\mu_i)=-\mu_i \bm{e}_i^T(Q-\mu_iI_S)^{-1},
\end{align}
Now, the optimal rate allocation policy for \eqref{eqn:opt_uncons_lag} will be a \emph{simple policy} (stationary and deterministic) \cite{Ross_CSMDP} and can be found using the policy iteration algorithm given in Algorithm~\ref{alg:policy_iter}. Let us denote the optimal policy for the unconstrained problem by $\zeta^\gamma$ and its average sampling rate by $\omega^\gamma$.

\begin{remark}
    To find the transition probability $\mathcal{P}(i,\mu_i,j)$, one can also evaluate the integral $\mathcal{P}(i,\mu_i,j)=\int_0^\infty P_{ij}(t)\mu_ie^{-\mu_it}\dd{t}$. However, for ease of implementation, we use the absorbing probabilities.
\end{remark}

\begin{algorithm}[t]
    \caption{Policy Iteration Algorithm for an SMDP}
    \begin{algorithmic}
    \renewcommand{\algorithmicrequire}{\textbf{Input:}}
    \renewcommand{\algorithmicensure}{\textbf{Output:}}
        \Require  SMDP described by the tuple $(\mathcal{S},\mathcal{A},P,R,H)$
        \Ensure Policy $\zeta^\gamma$
        \State {\bf Step 1:} Start with an initial arbitrary policy $\zeta$ where $\zeta_s$ is action in state $s$. 
         \State{\bf Step 2: (Relative Value Determination)} For the present policy $\zeta$, solve the following linear system of equations for $s=1,2,\ldots,S$, for the relative values $V_s$ and the long-run average reward $\mathcal{R}^{\zeta}$ of the policy $\zeta$,
         \[
          \mathcal{R}^{\zeta} H(s,\zeta_s) + V_s = \mathcal{R}(s,\zeta_s) + \sum_{s'=1}^S \mathcal{P}(s,\zeta_s,s') V_{s'},    
         \]
         by setting $V_S=0$.
         \State{\bf Step 3: (Policy Improvement)} For each state $s=1,2,\ldots,S$, set the alternative action $a_s$ to,
        \[
        \underset{a}{\text{arg max}} \ \ {\mathcal{R}(s,a) + \sum_{s'=1}^S \mathcal{P}(s,a,s') V_{s'} - \mathcal{R}^{\zeta}\mathcal{H}(s,a)}   
        \]
        \State {\bf Step 4:} Update the policy $\zeta_s=a_s$ for $s=1,2,\ldots,S$
        \State {\bf Step 5:} Stop when the two successive policies are the same and set $\zeta^\gamma=\zeta$. Otherwise, go to Step 2.   
    \end{algorithmic}
    \label{alg:policy_iter}
\end{algorithm}

Next, we need to translate our results from the unconstrained problem to the constrained problem. Since our action space limits the feasible sampling rates to be positive, the SMDP will be irreducible under any simple policy. Additionally, the simple policy $\mu_i=\Omega/2$ satisfies the sampling rate constraint. Therefore, from \cite{Ross_CSMDP}, we have that the optimal rate allocation policy for the constrained problem will be attained at either $\gamma=0$ or at some $\gamma>0$ which satisfies $\omega^\gamma=\Omega$. More specifically, if there exists a $\zeta^0$ such that $w^0\leq\Omega$, then $\zeta^0$ is the optimal policy for our constrained problem. Otherwise, if there exists some $\gamma>0$ such that $\omega^\gamma=\Omega$, then the optimal policy for the unconstrained problem with that particular $\gamma$ is optimal for the constrained problem. Finally, if both these conditions fail, then there exists some $\gamma$ such that $\lim_{\epsilon\downarrow0} \omega^{\gamma+\epsilon}=\Omega_0<\Omega<\Omega^0=\lim_{\epsilon\uparrow0} \omega^{\gamma+\epsilon}$. The existence of such a $\gamma$ is  shown in \cite{Ross_MDP}. Moreover, \cite{Ross_MDP} shows that $w^\gamma$ is non-increasing in $\gamma$, and hence, a bisection search can be used to narrow down the $\gamma$. However, in this case, the optimal policy is not a  simple policy but rather a \emph{semi-simple policy} (SSP), which means that, there exists at most one state for which the optimal policy needs to be randomized between two actions, and for every other state, the action is deterministic.

\subsubsection{Semi-Simple Policy (SSP)}
To find the optimal SSP, let us define $\bar\zeta=\lim_{\epsilon\downarrow0}\zeta^{\gamma+\epsilon}$ and $\underline{\zeta}=\lim_{\epsilon\uparrow0}\zeta^{\gamma+\epsilon}$. Let $\bar\omega$ and $\underline{\omega}$ be the average sampling rates of the policies $\bar\zeta$ and $\underline{\zeta}$. Then, $\bar\omega=\Omega_0$ and $\underline{\omega}=\Omega^0$. Then, the optimal SSP can be obtained by randomizing between $\bar\zeta$ and $\underline{\zeta}$ as follows. Denote by $\zeta_s$, the action taken under policy $\zeta$ when in state $s$. Let $\tilde{\zeta}^k$ be a policy such that $\tilde{\zeta}^k_s=\bar{\zeta}_s$ for $s>k$ and $\tilde{\zeta}^k_s=\underline{\zeta}_s$ for $s\leq k$. Let $\tilde{\omega}^k$ be the average sampling rate of $\tilde{\zeta}^k$. Since $\tilde{\omega}^0=\Omega_0$ and $\tilde{\omega}^S=\Omega^0$, there exists a $k'$ such that  $\tilde{\omega}^{k'-1}\leq\Omega<\tilde{\omega}^{k'}$. Then, the optimal SSP $\zeta^*$ is given by $\zeta^*_s=\bar\zeta_s$ for $s>k'$, $\zeta^*_s=\underline{\zeta}_s$ for $s<k'$ and for $s=k'$, we randomize between $\bar\zeta_{k'}$ and $\underline{\zeta}_{k'}$ with some probability $p$, where $p$ is chosen such that average sampling rate of $\zeta^*$ is $\Omega$.  The algorithm to find the optimal policy for the constrained problem is given in Algorithm~\ref{alg:opt_semi_simple}.

\begin{algorithm}[t]
    \caption{Optimal policy for the constrained SMDP}
    \begin{algorithmic}
    \renewcommand{\algorithmicrequire}{\textbf{Input:}}
    \renewcommand{\algorithmicensure}{\textbf{Output:}}
        \Require  $\gamma_u,\gamma_l,\epsilon_1,\epsilon_2\in \mathds{R}^+$
        \Ensure Policy $\zeta^*$
        \If{$w^0\leq \Omega$}
            \State Set $\zeta^*=\zeta^0$ and terminate.
        \EndIf
       \While{$\gamma_u-\gamma_l>\epsilon_1$}
            \State $\gamma=\frac{(\gamma_u+\gamma_l)}{2}$
            \If{$\omega^\gamma=\Omega$}
                \State Set $\zeta^*=\zeta^\gamma$ and terminate.
            \EndIf
            \If{$\omega^\gamma>\Omega$}
                \State $\gamma_l=\gamma$
            \Else
                \State $\gamma_u=\gamma$
            \EndIf
        \EndWhile
        \If{$w^{\gamma_l}-w^{\gamma_u}<\epsilon_2$}
            \State Set $\zeta^*=\zeta^{\gamma_u}$ and terminate. \Comment{This is a simple policy}
        \Else
            \State Construct a SSP by randomizing between $\zeta^{\gamma_u}$ and $\zeta^{\gamma_l}$ and set it as $\zeta^*$.
        \EndIf
    \end{algorithmic}
    \label{alg:opt_semi_simple}
\end{algorithm}

The analytical expressions for MBF and the average sampling rate for a SSP is given in Theorem~\ref{thrm:semi_simple}. 

\begin{theorem}\label{thrm:semi_simple}
    Let $r$ be the state of the SSP, where we randomize between rates $\mu_{r,1}$ and $\mu_{r,2}$ with probability $p$ and for every other state $i\neq r$, we have a fixed sampling rate of $u_i$. Let $\bm\mu^{(r)}=\{\mu_1,\mu_2,\dots,(p\mu_{r,1}+(1-p)\mu_{r,2},\dots,\mu_S\}$ be a rate allocation scheme and let $\tilde{\pi}_i^{(r)}$ be the proportion of time the ME was in state $i$ under the rate allocation scheme $\bm\mu^{(r)}$. Then, the MBF under the SSP, denoted by $\e[\tilde{\Delta}^r]$, is given by,
    \begin{align}
        \e[\tilde{\Delta}^r]=\frac{\mu_{r,1}\mu_{r,2}\left((p\mu_{r,1}+(1-p)\mu_{r,2})\tilde{\pi}_r^{(r)}(p\e[\tilde{F}_{i,\mu_{r,1}}]+(1-p)\e[\tilde{F}_{i,\mu_{r,2}}])+\sum_{i\neq r}^S\mu_i\tilde{\pi}_i^{(r)}\e[\tilde{F}_{i,\mu_i}]\right)}{(1-\tilde{\pi}_i)\mu_{r,1}\mu_{r,2}+\tilde{\pi}_i(p\mu_{r,1}+(1-p)\mu_{r,2})(p\mu_{r,2}+(1-p)\mu_{r,1})},\label{eqn:fresh_semi_simple}
    \end{align}
    and the sampling rate is given by, $\frac{\mu_{r,1}\mu_{r,2}\sum_{i=1}^S\tilde{\pi}_i^{(r)}\mu^{(r)}_i}{(1-\tilde{\pi}_i)\mu_{r,1}\mu_{r,2}+\tilde{\pi}_i(p\mu_{r,1}+(1-p)\mu_{r,2})(p\mu_{r,2}+(1-p)\mu_{r,1})}$.
\end{theorem}

The proof of Theorem~\ref{thrm:semi_simple} is given in Appendix~\ \ref{appen:thrm_semi_simple}.

\begin{remark}
    Our numerical results show that the optimal policy most likely turns out to be a simple policy, whereas semi-simple policies may sometimes arise as an artifact of the discretization of the action space when finding the optimal actions in the policy iteration algorithm. In fact, when we further refined the action space, we noted that our algorithm resulted in simple policies.
\end{remark}

\section{Optimum Monitoring of Multiple CTMCs} \label{sec:mult_ctmcs}
In this section, we consider the problem of monitoring multiple CTMCs simultaneously under a common average sampling budget. For simplicity, we will assume that each CTMC employs a single fixed sampling rate for all states (i.e., we employ a state-independent sampling policy for each CTMC). Now, the problem of interest is how we can utilize our sampling budget to monitor each CTMC so as to maximize the weighted sum of MBF across all CTMCs. 

Let $\tilde{\mu}_i$ denote the sampling rate allocated to the $i$th CTMC and let $\e[\Delta^{(i)}]$ denote its MBF. Then, the optimization problem  at hand can be expressed as follows,
\begin{mini}
    {\tilde{\rho}_l\leq\tilde{\mu}_i\leq \tilde{\rho}_u}{\sum_{i=1}^{C}w_i\e[\Delta^{(i)}]}
    {\label{eqn:het_ctmc}}
    {}
    \addConstraint{\sum_{i=1}^{C} \tilde{\mu}_i}{\leq \tilde{\Omega}},
\end{mini}
where $C$ is the number of different CTMCs, $\tilde{\Omega}$ is the total sampling budget, $\tilde{\rho}_u$ is the maximum feasible sampling rate, $\tilde{\rho}_l>0$ is chosen very close to zero, and $w_i$ is the weight associated with the $i$th CTMC, such that $\sum_{i=1}^\mathcal{C}w_i=1$. 

To solve the above problem, we will use a Lagrangian formulation following the techniques highlighted in  \cite{Ross_MDP, Ross_CSMDP}. Let $\tilde{\bm\mu}=\{\tilde{\mu}_1,\tilde{\mu}_2,\dots,\tilde{\mu}_{C}\}$ be the rate allocation vector of the CTMCs and define $L^\theta(\tilde{\bm\mu})=F(\tilde{\bm\mu})-\theta J(\tilde{\bm\mu})$ where $\theta\geq0$ is a Lagrangian multiplier, $F(\tilde{\bm\mu})$ is the weighted MBF and $J(\tilde{\bm\mu})$ is the sum of sampling rates under the rate allocation vector $\tilde{\bm\mu}$. Let $\tilde{\bm\mu}^\theta$ be the optimal rate allocation vector that maximizes $L^\theta(\tilde{\bm\mu})$. Since $L^\theta(\tilde{\bm\mu})$ is separable with respect to $\tilde{\mu}_i$, $\tilde{\mu}^{\theta}_i$ is given by $\tilde{\mu}^{\theta}_i =\argmax_{\tilde{\mu}_i\in[\tilde{\rho}_l, \tilde{\rho}_u]} w_i\e[\Delta^{(i)}]- \theta\tilde{\mu}_i$. 

The next lemma reveals several intriguing  properties of the above functions, some of which may be of separate interest for non-convex optimization problems.

\begin{lemma}\label{lem:continuity}
    $L^{\theta}(\tilde{\bm\mu}_\theta), F^{\theta}(\tilde{\bm\mu}_\theta)$ and $J^{\theta}(\tilde{\bm\mu}_\theta)$ are monotonically decreasing functions in $\theta$. $L^{\theta}(\tilde{\bm\mu}_\theta)$ is continuous with respect to $\theta$.  For $\tilde{\mu}_i^\theta$, if we always choose the smallest maximizer, then $J(\tilde{\bm\mu}_\theta)$ is right-continuous and if we always choose the largest maximizer then $J(\tilde{\bm\mu}_\theta)$ is left-continuous, with respect to $\theta$. If $\tilde{\mu}_i^\theta$ is unique for all $\theta$, then $J(\tilde{\bm\mu}_\theta)$ is continuous with respect to $\theta$.
\end{lemma}

The proof of Lemma~\ref{lem:continuity} is given in Appendix~\ \ref{appen:lem:continuity}. Now, to find the optimal rate allocation vector, we have that, if $J(\tilde{\bm\mu}_0)\leq\tilde{\Omega}$, then  $\tilde{\bm\mu}_0$ is the optimal rate allocation vector. If not, suppose for some $\theta^*>0$, we have $J(\tilde{\bm\mu}_{\theta^*})=\tilde{\Omega}$. Then, we have that,
\begin{align}
    L^{\theta^*}(\tilde{\bm\mu}_{\theta^*})=F(\tilde{\bm\mu}_{\theta^*})-\theta^* \tilde{\Omega}\geq L^{\theta^*}(\tilde{\bm\mu})=F(\tilde{\bm\mu})-{\theta^*} J(\tilde{\bm\mu}).
\end{align}
Since for any feasible $\tilde{\bm\mu}$, $J(\tilde{\bm\mu})\leq \Omega$, we have that $F(\tilde{\bm\mu}_\theta)\geq F(\tilde{\bm\mu})$. Further, from Lemma~\ref{lem:continuity}, we have that $J(\tilde{\bm\mu}_{\theta})$ is a monotonically decreasing function. Hence, if $\theta^*$ exists, it can be narrowed down using a bisection search since monotone functions have only jump discontinuities. If $\tilde{\mu}_i^\theta$ is  unique in the feasible region, then we have that a $\theta^*$ always exists. In fact, the uniqueness of $\tilde{\mu}_i^\theta$ at $\theta^*$ is enough to guarantee the convergence to an optimal policy. However, if no such $\theta$ exists, since the constraints are linear, the problem will still satisfy the linear constraint qualification (LCQ) condition. Therefore, the KKT conditions give the necessary conditions for optimality. As an artifact of the KKT conditions, we have that, at the optimal point, the sum constraint must be satisfied if $J(\tilde{\bm\mu}_{0})>\tilde{\Omega}$. Thus, in this case we can find a locally optimal solution using a projected gradient decent where the projection is done on to a scaled simplex\cite{Markov_machines}. Thus, our strategy to find the optimal rates is two folds. First, we use a bisection search to narrow down  $\theta^*$ for which $J(\tilde{\bm\mu}_{\theta^*})=\tilde{\Omega}$. The bisection search will yield a decreasing sequence $\bar\theta_n$ which converges to $\theta^*$. If $J(\tilde{\bm\mu}_{\theta})$ is continuous at $\theta^*$, then  the $\lim_{n\to \infty}\tilde{\bm\mu}_{\bar \theta_n}$ will yield the optimal policy. Otherwise, we find all the maximizers of $L^{\theta^*}(\tilde{\bm\mu})$. If one of them has a $J(\tilde{\bm\mu}_{\theta^*})=\tilde{\Omega}$, then that is optimal. If not, we find locally optimal $\tilde{\bm\mu}_{\theta}$ values using a projected gradient descent and select the best out of the gradient descent and the bisection search. The algorithm to find the optimal sampling rates is given in Algorithm~\ref{alg:opt_rate_mm}.

\begin{algorithm}[t]
    \caption{Optimal rate allocation policy for multiple CTMCs}
    \begin{algorithmic}
    \renewcommand{\algorithmicrequire}{\textbf{Input:}}
    \renewcommand{\algorithmicensure}{\textbf{Output:}}
        \Require  $\theta_u,\theta_l,\epsilon_1,\epsilon_2\in \mathds{R}^+$
        \Ensure The optimal rate vector $\tilde{\bm\mu}^*$
        \If{$J(\tilde{\bm\mu}_0)\leq \tilde{\Omega}$}
            \State Set $\tilde{\bm\mu}^*=\tilde{\bm\mu}_0$ and terminate.
        \EndIf
       \While{$\theta_u-\theta_l>\epsilon_1$}
            \State $\theta=\frac{(\theta_u+\theta_l)}{2}$
            \If{$J(\tilde{\bm\mu}_\theta)=\tilde{\Omega}$}
                \State Set $\tilde{\bm\mu}^*=\tilde{\bm\mu}_\theta$ and terminate.
            \EndIf
            \If{$J(\tilde{\bm\mu}_\theta)>\tilde{\Omega}$}
                \State $\theta_l=\theta$
            \Else
                \State $\theta_u=\theta$
            \EndIf
        \EndWhile
        \If{$J(\tilde{\bm\mu}_{\theta_l})-J(\tilde{\bm\mu}_{\theta_u})<\epsilon_2$}
            \State Set $\tilde{\bm\mu}^*=\tilde{\bm\mu}_{\theta_l}$ and terminate. 
        \Else
            \State Do a projected gradient descent to find a locally optimal solution. Set $\tilde{\bm\mu}^*$ as the best out of the bisection search (all possible solutions at $\theta$) and the projected gradient descent.
        \EndIf
    \end{algorithmic}
    \label{alg:opt_rate_mm}
\end{algorithm}

\begin{remark}
   Algorithm \ref{alg:opt_rate_mm} can be used to find the optimal rate allocation scheme for monitoring a set of general CTMCs with any estimator of choice as long as the freshness expressions can be computed efficiently.
\end{remark}

\section{Numerical Results} \label{sec:numer}
In this section, we comprehensively evaluate the performances of the proposed structured estimators and compare our rate allocation policies with suitable benchmarks. First, we will compare the MBF of our simple two-stage estimators and then move on to compare the MBF of the proposed multi-stage estimators. Next, we will compare several state-dependent sampling  policies for monitoring a single CTMC. Finally, we will compare different rate allocation policies for monitoring multiple heterogeneous CTMCs. For reproducibility purposes of the results, where possible, we have included detailed descriptions of the generator matrices used in each experiment in Table \ref{tab:gen_mats}.

\begin{remark}
    Even though most of our theoretical results and algorithms hold for generic CTMCs, we have preferred to use time-reversible CTMCs in the numerical examples. The underlying reason is that we have closed-form expressions for time-reversible CTMCs for which Algorithms~\ref{alg:opt_semi_simple} and \ref{alg:opt_rate_mm} can be applied without any additional computational overhead which would stem from the numerical integration in \eqref{eqn:Fint}.
\end{remark}

\begin{figure}[t]
    \centering
    \includegraphics[width=0.5\textwidth]{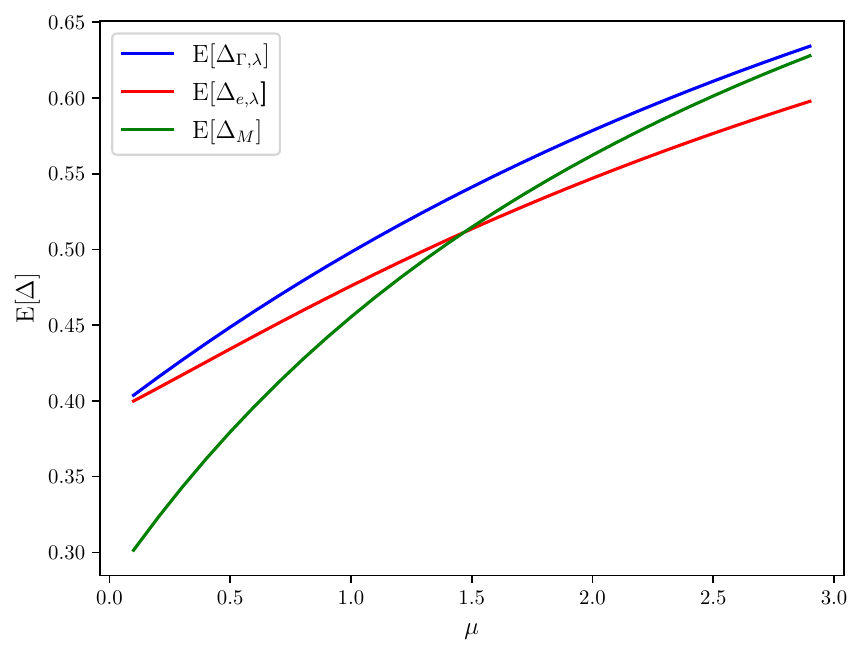}
    \caption{Variation of MBF  with the sampling rate for $\Gamma=10$ and $\lambda=\frac{1}{\tau^*}$ for a generic 4-state CTMC.}
    \label{fig:var_mu}
\end{figure}

To evaluate the performance of the simple  two-stage estimators, we will be considering CTMCs with a unique stationary maximum. Moreover, as the example for a time-reversible CTMC, we will be using a star topology, where the central state (i.e., central node) connects to every other state and no connections are present between any other states, similar to Fig.~\ref{fig:tr_examples}(a). This structure is naturally a time-reversible CTMC, and the rates are set so that the central server has the highest stationary probability. First, we will compare how the MBF varies under the martingale, exponential and Erlang estimators, as we vary the sampling rate, for a generic CTMC with four states. As illustrated in Fig.~\ref{fig:var_mu}, the EXPE (red) outperforms the ME (green) at low sampling rates, while the opposite behavior is observed at high sampling rates. This is due to the fact that, for high sampling rates, when the exponential timer runs out before we take the next sample, it would most probably run out at smaller time intervals (smaller than $\tau^*$). Thus, our estimator will shift to $i^*$ prematurely reducing the MBF metric. At all sampling rates, we can see that the ERLE (blue) outperforms the other two estimators. For higher sampling rates, the performance of the ME approaches that of the ERLE, while for lower sampling rates the EXPE exhibits comparable performance to the ERLE.

\begin{figure}[t]
    \centering
    \includegraphics[width=0.5\textwidth]{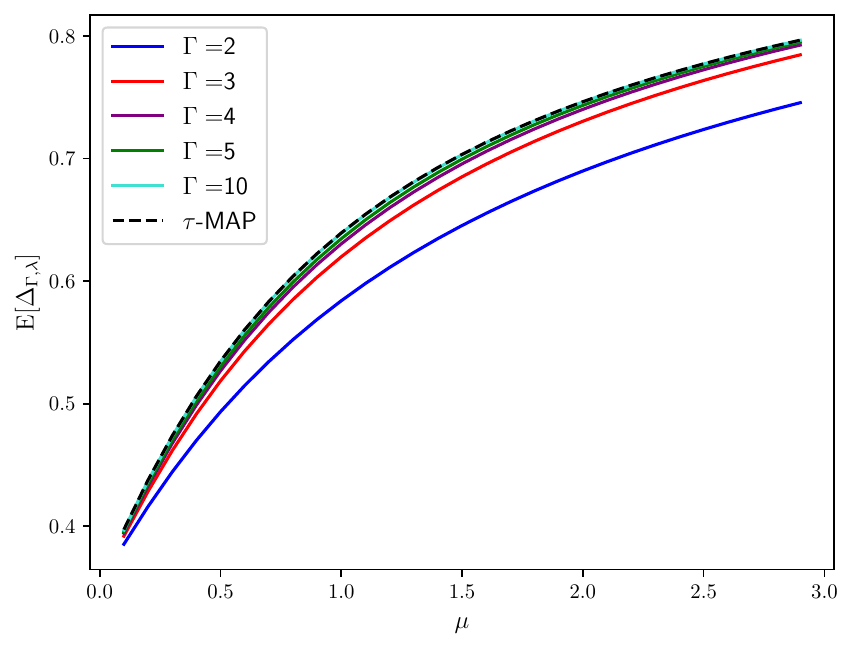}
    \caption{Variation of MBF with the sampling rate for different $\Gamma$ values with $\lambda=\frac{1}{\tau^*}$ and $\tau=\tau^*$ for a time-reversible 5-state CTMC.}
    \label{fig:var_gamma_mu}
\end{figure}

Next, we evaluate how the MBF varies as we increase the $\Gamma$ parameter of the ERLE for a time-reversible CTMC with $5$ states. As depicted in Fig.~\ref{fig:var_gamma_mu}, higher $\Gamma$ values lead to better MBF. Moreover, the curves converge and approach the $\tau$-MAP curve as $\Gamma$ increases. As evident from Fig.~\ref{fig:var_mu} and Fig.~\ref{fig:var_gamma_mu}, $\tau$-MAP estimators clearly exhibit better MBF performance compared to exponential and Erlang estimators. Hence, going forward, we will only use the $\tau$-MAP estimator as the representative for our simple two-stage estimators. 

\begin{figure}[t]
    \centering
    \begin{subfigure}[b]{0.45\textwidth}
        \centering
        \includegraphics[width=\textwidth]{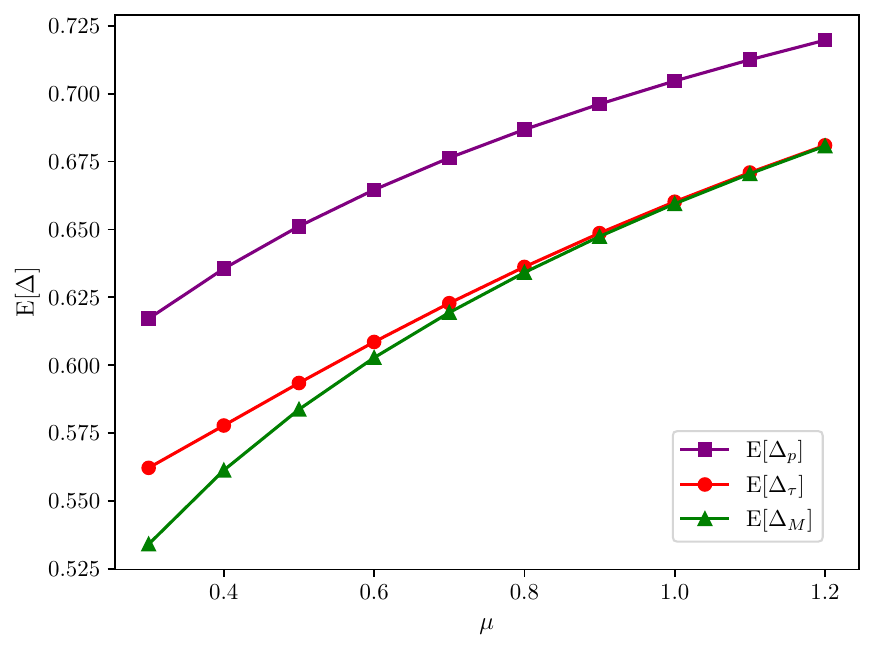}
        \subcaption{Finite BDC with $S=4$.}
        \label{fig:var_omega_0}
    \end{subfigure}
    \hfill 
    \begin{subfigure}[b]{0.45\textwidth}
        \centering
        \includegraphics[width=\textwidth]{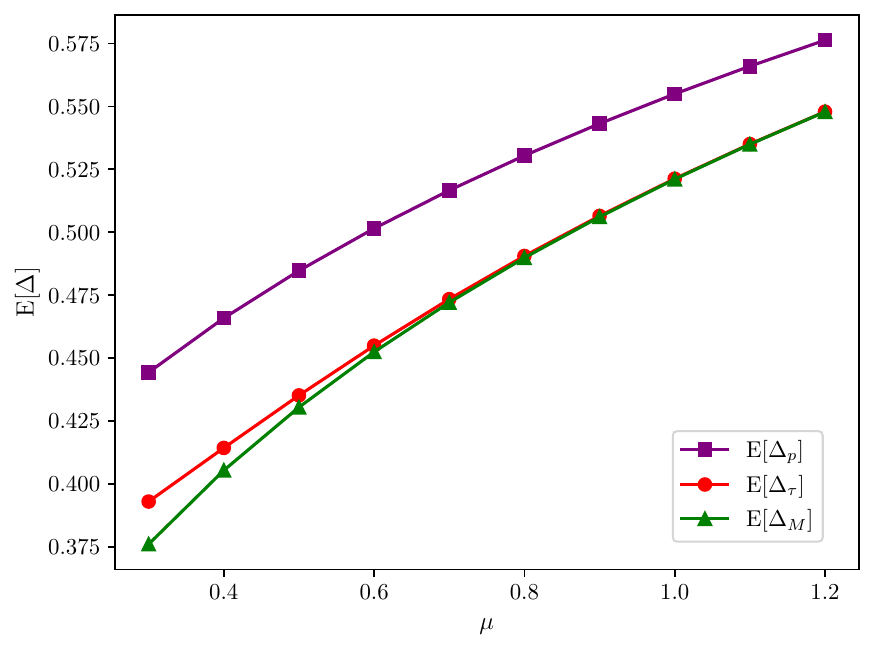}
        \subcaption{Finite BDC with $S=6$.}
        \label{fig:var_omega_0_2}
    \end{subfigure}
\par\bigskip\smallskip
    \begin{subfigure}[b]{0.45\textwidth}
        \centering
        \includegraphics[width=\textwidth]{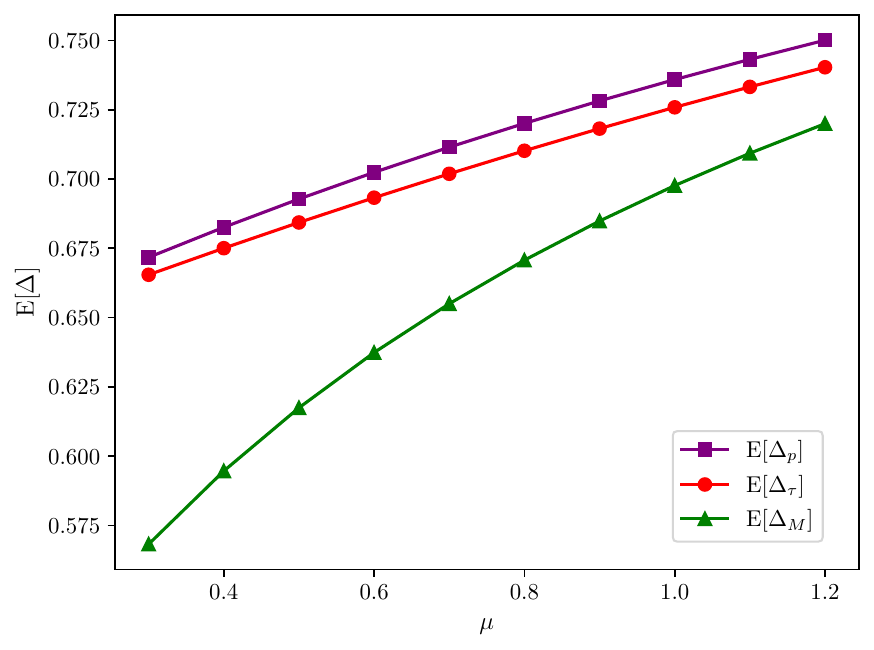}
        \caption{Star topology with $S=4$.}
        \label{fig:var_omega_0_star_4}
    \end{subfigure}
    \hfill 
    \begin{subfigure}[b]{0.45\textwidth}
        \centering
        \includegraphics[width=\textwidth]{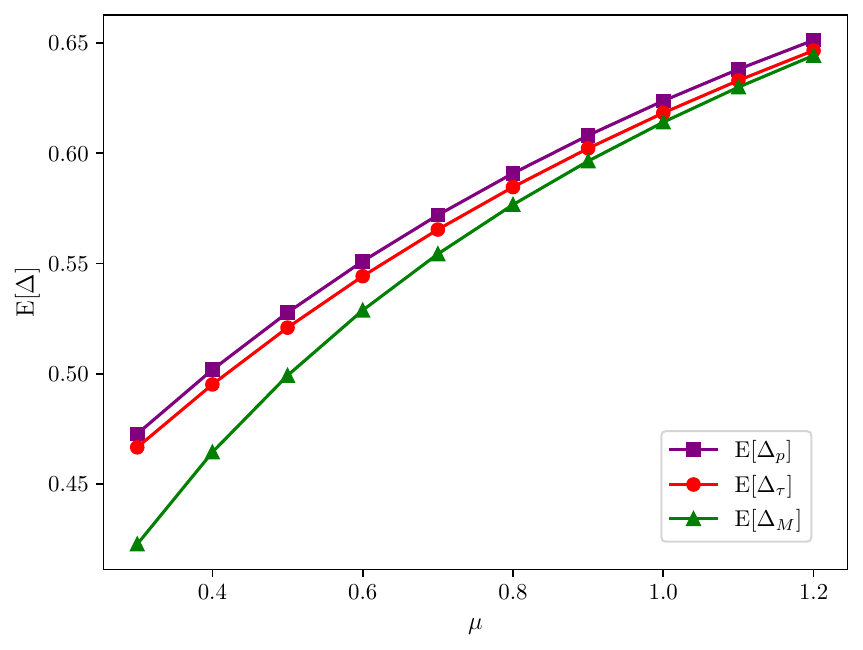}
        \caption{Star topology with $S=6$.}
        \label{fig:var_omega_0_star_6}
    \end{subfigure}
    \caption{Variation of MBF with the sampling rate $\mu$ for different CTMCs with a unique stationary maximum.}
    \label{fig:var_mu_multi}
\end{figure}

Now, we compare the performance of our multi-stage estimators with that of our simple two-stage estimators and evaluate how various state-dependent sampling policies affect the MBF. In here, for the $p$-MAP estimator, we always assume that $\tau_{i,k}=\tau^*_{i,k}$ unless otherwise specified, and for the $\tau$-MAP estimator, $\tau=\tau^*$ is used. To construct time-reversible CTMCs, in addition to the star topology, we also use  finite birth death chains (BDC) which are well-known to be time-reversible. We will sample these CTMCs at a fixed state-independent sampling rate $\mu$ and will evaluate the MBF for the martingale, $\tau$-MAP and $p$-MAP estimators as we vary the sampling rate $\mu$. As evidenced by Fig.~\ref{fig:var_mu_multi}, there is a significant gain ($\sim 17\%$ at  $\mu=0.3$ in Fig.~\ref{fig:var_mu_multi}(a)) attained in terms of MBF as a result of replacing the ME with the $p$-MAP estimator. Moreover, we observe that the $p$-MAP estimator exhibits considerable freshness gains ($\sim10\%$ at  $\mu=0.3$ in Fig.~\ref{fig:var_mu_multi}(a)) over the $\tau$-MAP estimator by replacing the single threshold in the $\tau$-MAP estimator with multiple thresholds. 
 
In the next experiment, we compare how different state-dependent sampling policies affect the MBF. In here, we use finite-state BDCs to evaluate the performance our state-dependent sampling policies. We denote by $\Delta_M^*$, $\Delta_\tau^*$ and $\Delta_p^*$, the MBF of the three estimators under their optimal sampling rates computed using the SMDP framework. We compare these optimal state-dependent sampling policies against a uniform sampling policy, which allocates $\mu_i=\Omega$ for all $i$. This rate allocation policy is trivially contained within the feasible set of our original optimization problem for state-dependent sampling policies, and hence, it provides a natural lower bound for the MBF curves under the optimal policy. Let us denote the MBF of the three estimators under the uniform sampling policy by $\Delta_M^U$, $\Delta_\tau^U$ and $\Delta_p^U$.

\begin{figure}[t]
    \centering
    \includegraphics[width=0.5\textwidth]{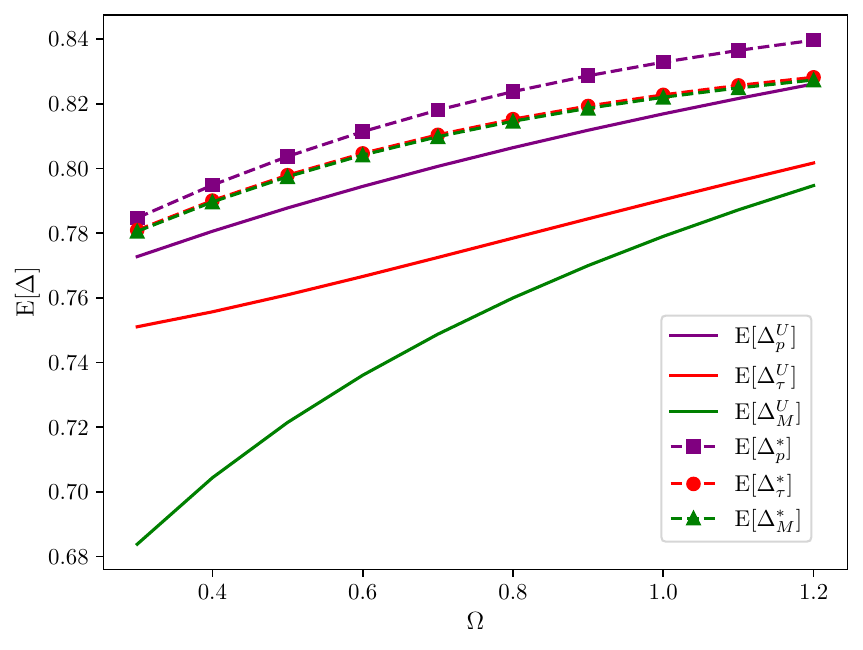}
    \caption{Variation of MBF with the sampling budget $\Omega$, under different state-dependent sampling policies, for a finite BDC ($S=4$) with a unique stationary maximum.}
    \label{fig:var_omega}
\end{figure}

Fig.~\ref{fig:var_omega} illustrates the variation of MBF of the three estimators under the considered sampling policies for a BDC with a unique stationary maximum. As illustrated, under any given rate allocation policy, the $p$-MAP estimator is always superior to the rest whereas $\tau$-MAP estimator is second to the $p$-MAP estimator. Moreover, we see that under the optimal rate allocation policies, a significant improvement ($\sim15\%$ for ME and $\sim4\%$ for $\tau$-MAP, at $\mu=0.3$) can be observed compared to the uniform policy, highlighting the importance of a state-dependent rate allocation scheme. Also, note that, when the optimal state-dependent sampling policy is employed, the performance gap among the structured estimators narrows. This is mainly due to the fact that, the optimal sampling policy will try to allocate higher rates to less probable states, and low rates to more probable states. Therefore, when in a less probable state, we will quickly sample and acquire a fresher update, whereas in more probable states, we are less likely to obtain new samples for longer durations. This improves the overall MBF of the system by balancing out one of the main drawbacks of the ME, that is, having long sojourn times in less probable states.

Next, we compare how MBF varies with the number of stages $K_i$ used in the $p$-MAP estimator to approximate an infinitely oscillating MAP estimator. Here, we use a non-time-reversible ring structure similar to that of Fig.~\ref{fig:map_osc}, to construct a CTMC with an infinitely oscillating MAP estimator. For this, we consider a CTMC with $4$ states placed in a ring structure, where the transition rates alternate between two values. In particular, we transition from state $1$ to $2$ and from state $3$ to $4$ at a rate of $1$, whereas a rate of $0.75$ is used for transitions between states $2$ to $3$ and from state $4$ to $1$. No other transitions exist. Further, we assume a fixed state-independent sampling rate for this problem. Since the MAP estimator is infinitely oscillating, it will have infinitely many transition points$\tau_{i,k}^*$ given a starting state $i$. Ideally, if we use all these infinitely many transition points, our $p$-MAP estimator would resemble the MAP estimator exactly. In here, we will be using a subset of these optimal transition points for our $p$-MAP estimator. To construct these subsets we use two main methods. In the first method, for each starting state, we use the first $K_i$ transition points of the MAP estimator to construct the $p$-MAP estimator, and in the second method, we periodically sample the first nine transition points of the MAP estimator at different periods leading to different $K_i$ values for our $p$-MAP estimator. 

\begin{figure}[t]
    \centering
    \begin{subfigure}[b]{0.45\textwidth}
        \centering
        \includegraphics[width=\textwidth]{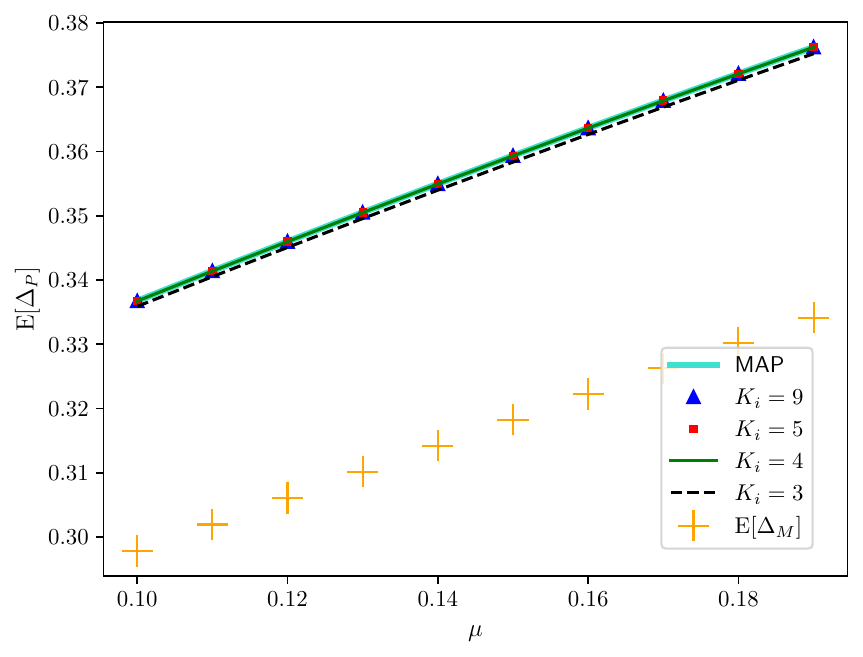}
        \caption{Using the first $K_i$ optimal transition points.}
        \label{fig:p_map_var}
    \end{subfigure}
    \hfill 
    \begin{subfigure}[b]{0.45\textwidth}
        \centering
        \includegraphics[width=\textwidth]{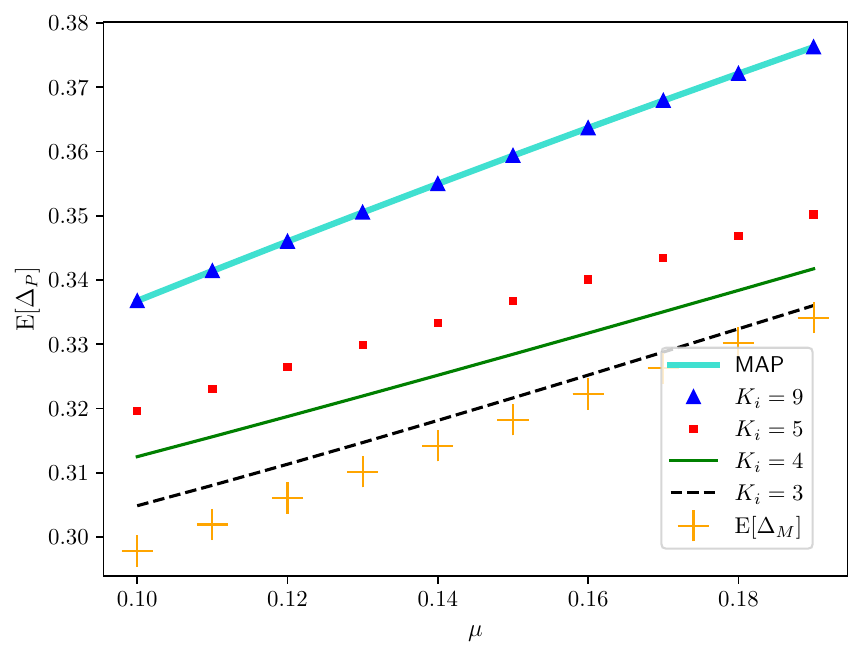}
        \caption{Periodically sampling the first nine optimal transition points.}
        \label{fig:p_map_var_periodic}
    \end{subfigure}
    \caption{Variation of MBF of the $p$-MAP estimator with the number of intermediate stages $K_i$ used to approximate an infinitely oscillating MAP estimator.}
    \label{fig:var_K_i}
\end{figure}

Fig.~\ref{fig:var_K_i}(a) and Fig.~\ref{fig:var_K_i}(b) illustrate the MBF curves under these two methods, respectively. As seen, in both cases, as the number of stages $K_i$ increases, the $p$-MAP estimators approach the MAP estimator. Further, we see that, having an exact approximation of the MAP estimator at the initial transition points is more important than the latter transition points. In other words, from a freshness perspective, it is sufficient, if our estimators approximate the MAP estimator closely when the age of the estimator $\delta(t)$ is low, and approximate it more coarsely when the age of the estimator $\delta(t)$ is high.

\begin{figure}[t]
    \centering
    \includegraphics[width=0.5\textwidth]{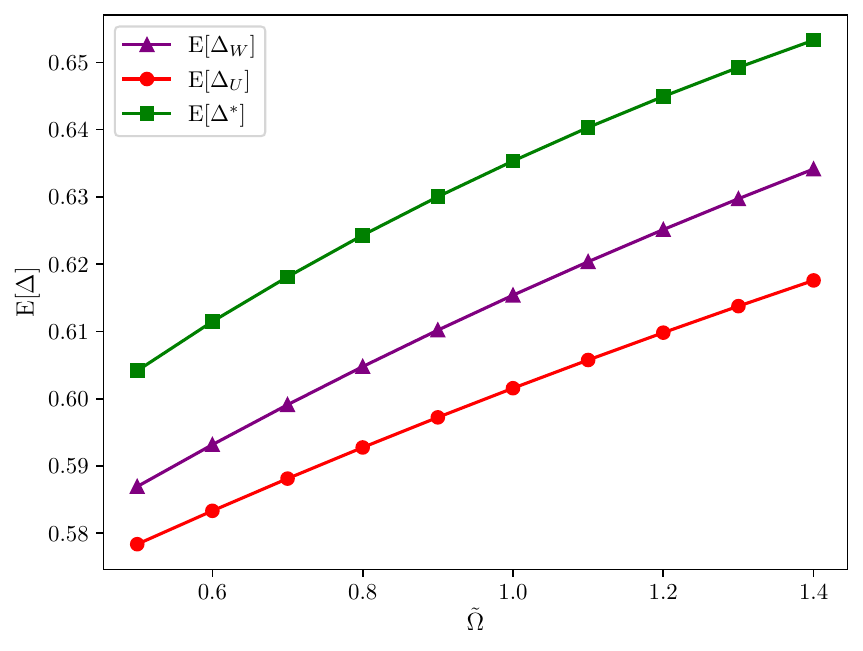}
    \caption{Variation of MBF with respect to the sampling budget $\tilde{\Omega}$, under different rate allocation policies, for monitoring finite BDCs ($C=5$) using the $p$-MAP estimator.}
    \label{fig:fresh_mm}
\end{figure}

Finally, we consider the problem of monitoring multiple heterogeneous CTMCs and evaluate how the MBF varies for different rate allocation policies. Here, we only consider the $p$-MAP estimator with optimal transition points (i.e., $\tau_{i,k}=\tau^*_{i,k}$) and for the heterogeneous CTMCs again we use finite BDCs. Here, we compare the optimal rate allocation policy against two benchmark policies. Once again we have a uniform policy which in this scenario assigns the same rate to all the CTMCs (i.e., $\tilde{\mu}_i=\frac{\tilde{\Omega}}{C}$) and then, we have a weight-based policy which assigns the rates based on the weights allocated for each CTMC (i.e., $\tilde{\mu}_i=w_i\tilde{\Omega}$). Let us denote the MBF under the optimal policy, uniform policy and the weight-based policy using $\e[\Delta^*]$, $\e[\Delta_U]$ and $\e[\Delta_W]$, respectively.

\begin{figure}[t]
    \centering
    \includegraphics[width=0.5\textwidth]{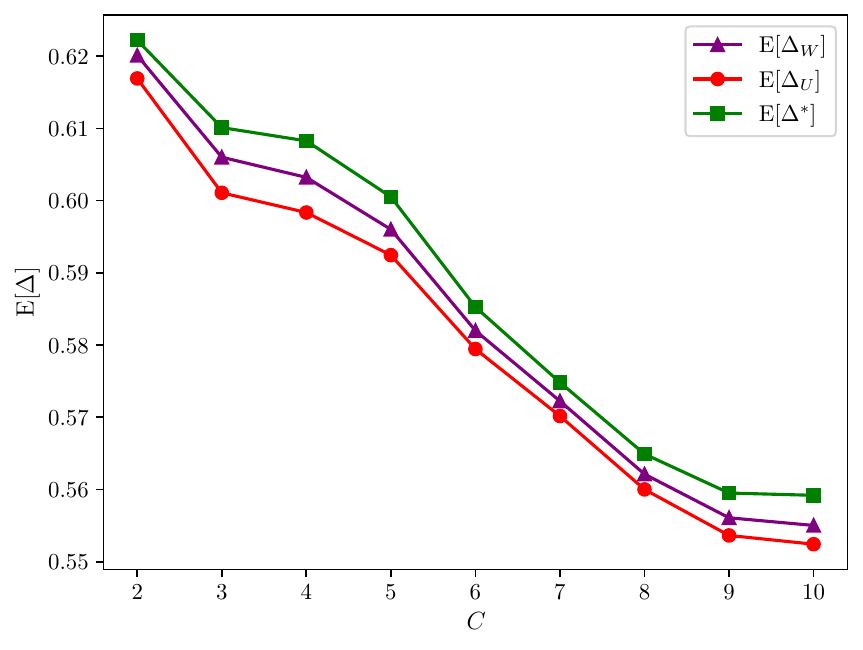}
    \caption{Variation of MBF with $C$ for a fixed sampling budget, when monitoring $C$ finite-state BDCs, under different rate allocation policies with $p$-MAP estimator. Here, $\tilde{\Omega}=1$ with $w_i=\frac{2i}{c(c+1)}$ when $C=c$).}
    \label{fig:fresh_mm_sources}
\end{figure}

Fig.~\ref{fig:fresh_mm}, illustrates the variation of these three freshness metrics as we vary the total sampling budget $\tilde{\Omega}$. As seen, the optimal curve outperforms the rest by a significant margin, whereas the weight-based policy slightly outperforms the uniform policy. The same trend was observed, when we varied the number of CTMCs monitored with the sampling budget fixed (see Fig.~\ref{fig:fresh_mm_sources}).  Further, we also note that our algorithm never resulted in a projected gradient descent for the selected tolerance values ($\epsilon_1=10^{-5},\epsilon_2=10^{-3})$, and the bisection search always gave the optimal solution. In here, we find the optimal $\tilde{\mu}_i^\theta$ by discretizing and doing an exhaustive search in the interval $[\tilde{\rho}_l,\tilde{\rho}_u]$. Therefore, we also observed that our algorithm would always meet the required tolerance bounds by further refining the search interval for smaller tolerance values. 

Finally, we conclude our numerical results by providing the generator matrices used in each of our single-source experiments in  Table~\ref{tab:gen_mats} for reproducibility of the graphs.

\begin{table}[h!]
    \centering
    \caption{CTMCs used in the experiments.}
    \label{tab:gen_mats}
    \begin{tabular}{|c|c|c|}
        \hline 
        Figure & Topology & Generator matrix \\
        \hline \hline 
        Fig.~\ref{fig:var_mu} &  4 state generic & \small{$\begin{bmatrix}
            -1.70&0.41&0.53&0.76\\
            1.03&-2.17&0.83&0.31\\
            1.11&0.78&-2.74&0.85\\
            1.15&1.13&0.71&-2.99
        \end{bmatrix}$ }\\
        \hline 
          Fig.~\ref{fig:var_gamma_mu} & 5 state star &  \small{$\begin{bmatrix}
            -3.99&0.77&0.69&1.29&1.24\\
            0.70&-0.70&0&0&0\\
            0.84&0&-0.84&0&0\\
            0.71&0&0&-0.71&0\\
            0.48&0&0&0&-0.48
        \end{bmatrix}$} \\
        \hline 
        Fig.~\ref{fig:var_mu_multi}(a) &  4 state BDC & \small{$\begin{bmatrix}
            -0.79&0.79&0&0\\
            1.71&-1.97&0.26&0\\
            0&1.08&-2.73&1.65\\
            0&0&0.62&-0.62
        \end{bmatrix}$} \\
        \hline
        Fig.~\ref{fig:var_mu_multi}(b) & 6 state BDC &\small{$\begin{bmatrix}
            -0.62&0.62&0&0&0&0\\
            0.63&-1.58&0.95&0&0&0\\
            0&1.748&-3.55&1.81&0&0\\
            0&0&0.47&-2.22&1.75&0\\
            0&0&0&1.15&-1.88&0.73\\
            0&0&0&0&1.91&-1.91
        \end{bmatrix}$}\\
        \hline 
         Fig.~\ref{fig:var_mu_multi}(c) & 4 state star & \small{ $\begin{bmatrix}
            -2.41&1.25&0.50&0.66\\
            0.36&-0.36&0&0\\
            1.20&0&-1.20&0\\
            1.18&0&0&-1.18
        \end{bmatrix}$} \\
        \hline 
         Fig.~\ref{fig:var_mu_multi}(d) & 6 state star &\small{$\begin{bmatrix}
            -4.83&1.24&0.83&1.25&0.88&0.63\\
            1.08&-1.08&0&0&0&0\\
            0.31&0&-0.31&0&0&0\\
            0.99&0&0&-0.99&0&0\\
            1.01&0&0&0&-1.01&0\\
            1.17&0&0&0&0&-1.17
        \end{bmatrix}$}\\
        \hline 
         Fig.~\ref{fig:var_omega} &  4 state BDC & \small{$\begin{bmatrix}
            -1.02&1.02&0&0\\
            1.05&-2.21&1.16&0\\
            0&0.61&-2.18&1.57\\
            0&0&0.26&-0.26
        \end{bmatrix}$ }\\
        \hline
    \end{tabular}
\end{table}

\section{Conclusion}
In this work, we have moved away from the norm so far in the literature, and have taken upon the challenge of modeling information freshness for non-martingale estimators. In particular, we have introduced several estimators that can enhance freshness by closely modeling MAP estimators. In particular, we have shown that the $p$-MAP estimator can exactly model the MAP estimator for time-reversible CTMCs, and can approximate the MAP estimator of non-time-reversible CTMCs when we increase the number of intermediate stages. We have also presented an algorithm based on an SMDP formulation to find the optimal state-dependent sampling policy for the case of a single monitored CTMC process. Using $p$-MAP estimators and state-dependent sampling policies, replacing the conventional MEs and state-oblivious sampling policies, respectively, we have shown substantially enhanced MBF (up to $17\%$ improvement) in the scenarios we studied. We also derived an optimal rate allocation policy when multiple heterogeneous CTMCs are monitored using state-oblivious sampling, with various estimators. In this case as well, with optimal sampling rate allocation, the MBF performance of the system is significantly improved, in comparison with two benchmark policies. Future directions in this line of work involve the analysis of MBF in settings where status feedback from queries are delayed, and utilization of the estimators for scenarios beyond query-based sampling. We also aim to extend these estimators to the study of age-oriented freshness metrics such as the AoII.

\appendices

\section{Proof of Lemma~\ref{lem:T}} \label{appen:lem_T}
Assume the CTMC has a unique stationary maximum. Let $\tilde{i}=\inf\{\argmax_{i\in \mathcal{S},i\neq i^*}\pi_i\}$ and $\epsilon=\frac{\pi_{i^*}-\pi_{\tilde{i}}}{4}>0$. Since $X(t)$ is ergodic, we have that $\lim_{t\to\infty}P(t)=\mathbf{1}\bm{\pi}^T$ which is equivalent to $\lim_{t\to\infty}\|P(t)-\mathbf{1}\bm{\pi}^T\|_{F}=0$ where $\|\cdot\|_F$ is the Frobenius norm. Therefore, $\exists~\tau^*>0$ such that $\forall t>\tau^*$, $\|P(t)-\mathbf{1}\bm{\pi}^T\|_{F}<\epsilon$. Now, for the rest of the proof, consider that $t>\tau^*$. This implies that $|P_{ij}(t)-\pi_j|<\epsilon$. Then, for $j\neq i^*$, the following relations hold,
\begin{align}
    P_{ii^*}-P_{ij}&>(\pi_{i^*}-\epsilon)-(\pi_{j}+\epsilon)
    >\pi_{i^*}-\pi_{\tilde{i}}-2\epsilon=\frac{\pi_{i^*}-\pi_{\tilde{i}}}{2}>0.
\end{align}
Therefore, for $t>\tau^*$, we have that $\argmax_{i\in \mathcal{S}}\bm{e}_j^TP(t)=i^*$.

\section{Proof of Lemma~\ref{lem:Pij}} \label{appen:lem:Pij}
Recalling $\bm{e}_i$ to be the vector of all zeros except for a one at the $i$th index, the $P_{ij}(t)$ for a time-reversible CTMC can be found as follows,
\begin{align}
    P_{ij}(t)&=\bm{e}_i^Te^{tQ}\bm{e}_j\\
    &=\bm{e}_i^T\left(I+Qt+\frac{Q^2t^2}{2!}+\dots\right)\bm{e}_j\\
    &=\bm{e}_i^T\Pi^{-\frac{1}{2}}U\left(I+Dt+\frac{D^2t^2}{2!}+\dots\right)\Pi^{\frac{1}{2}}U^T\bm{e}_j\\
    &=\bm{e}_i^T\Pi^{-\frac{1}{2}}U\begin{pmatrix}
    e^{-d_1t}&0&\dots&0\\
    0&e^{-d_2t} &\dots&0\\
    \vdots&\vdots&\ddots&\vdots\\
    0&0&0&e^{-d_St}
    \end{pmatrix}U^T\Pi^{\frac{1}{2}}\bm{e}_j.
\end{align}
Multiplying the above matrices yields the desired result.

\section{Proof of Lemma~\ref{lem:finite_roots}} \label{appen:lem:finite_roots}
Note that, from Lemma~\ref{lem:Pij}, we have that $P_{ij}(t)$ is a linear combination of exponential functions of the form $e^{-d_kt}$. Therefore, the difference $f^i_{jk}(t)=P_{ij}(t)-P_{ik}(t)$, for $j\neq k$, is also another linear combination of exponential functions. To proceed with the proof, we will first show that a function  of the form $g(t)=\sum_{m=1}^n\alpha_me^{\beta_mt}\neq0$, will have only finitely many roots. Here, when we use the term roots, we always refer to real-valued roots. Note that, if $g(t)$ is a constant, then the result is trivial (no roots). Now, assume $g(t)$ is not a constant (i.e., at least one $\beta_m\neq0$). Without loss of generality, assume $\beta_1\leq\beta_2\leq \dots\leq\beta_n$. Now, let us divide $g(t)$ by its smallest exponential component to obtain a new function $h(t)$ and denote this operation by $H$. Thus, we have $h(t)=H(g(t))=g(t)e^{-\beta_1t}$. Further, denote the derivative of a function  $f$ as $f^\prime$. Let $g_1(t)=h^\prime(t)$ and $h_1(t)=H(g_1(t))$ (i.e., $g_1(t)$ divided by its smallest exponential component). Now, similarly define $g_i(t)=h^\prime_{i-1}(t)$ and $h_i(t)=H(g_i(t))$. Note that, the operation $H$ followed by differentiation, at least removes one term from sum of exponentials, and hence, after finite iterations, we will have $h_i(t)=\bar\beta\neq0$ for some $\bar\beta\in \mathbb{R}$ (no roots). Now, by the virtue of Rolle's theorem, we have that the number of roots of  a continuous function $f$ is less than the number of roots of $f^\prime$ plus one. Also note that, the number of roots of $g_i(t)$ is one more than the number of roots of $h_i(t)$. Since $h_i(t)$ has no roots for some $i$, by repeated application of Rolle's theorem for $j<i$, we have that the number of roots of $g(t)$ is at most $2i+1$.

Therefore, we have that $f^i_{jk}(t)$ has only finitely many roots. Now, consider the ordered sequence of roots  $\{t_m\}_m$ generated by  these functions where $t_m\in\tilde{R}=\{t:\prod_{j<k}f^i_{jk}(t)=0\}$. For any $t\in (t_m,t_{m+1})$, we have that $f^i_{jk}(t)$ is either strictly positive or negative throughout the interval, and hence, the MAP estimate is constant in that interval. In particular, a subset of points in $\tilde{R}$ would define the transition points of the MAP estimate. Moreover, since $\tilde{R}$ has only finitely many points, the MAP estimate is piecewise constant with finitely many transition points.

\section{Proof of Theorem~\ref{thrm:erlang_BF}} \label{appen:erlang_BF}
Let $Z(t)=(X(t),\hat{X}_{\Gamma,\lambda}(t),Y(t))$. Then, $Z(t)$ is an irreducible finite-state Markov chain. Let $\bm{\Phi}=\{\phi_{i,j,k}\}_{i,j,k \in \mathcal{S}}$ be the stationary distribution of $Z(t)$. Then, the global balanced equations yield the following:
\begin{itemize}
    \item $1<k<\Gamma$
    \begin{align}
    (q_i+\mu+\lambda \Gamma)\phi_{i,j,k}=\lambda \Gamma\phi_{i,j,k-1}+\sum_{l\neq i}\phi_{l,j,k}q_{li}.
    \end{align}
    \item $i \neq j$, $k=1$
    \begin{align}
      (q_i+\mu+\lambda \Gamma)\phi_{i,j,1}=\sum_{l\neq i}\phi_{l,j,1}q_{li}.
    \end{align}
    \item  $i=j$, $k=1$
    \begin{align}
      (q_i+\lambda \Gamma)\phi_{i,i,1}=\sum_{l\neq i}\phi_{l,i,1}q_{li}+\mu\sum_{m>1}\sum_l\phi_{i,l,m}
      +\mu\sum_{l\neq i}\phi_{i,l,1}.
    \end{align}
    \item $k=\Gamma$
    \begin{align}
     (q_i+\mu)\phi_{i,i^*,\Gamma}=\lambda \Gamma \sum_{l}\phi_{i,l,\Gamma-1}+\sum_{l\neq i} \phi_{l,i^*,\Gamma}q_{li}.
    \end{align}
\end{itemize}
Rearrangement of the above equations along with the facts $\sum_{l,m}\phi_{i,l,m}=\pi_i$ and $q_{ii}=-q_i$, give the following set of equations:
\begin{itemize}
    \item $1<k<\Gamma$
    \begin{align}
    (\mu+\lambda \Gamma)\phi_{i,j,k}=\lambda \Gamma\phi_{i,j,k-1}+\sum_{l}\phi_{l,j,k}q_{li}.
    \end{align}
    \item $i \neq j$, $k=1$
    \begin{align}
      (\mu+\lambda \Gamma)\phi_{i,j,1}=\sum_{l}\phi_{l,j,1}q_{li}.
    \end{align}
    \item  $i=j$, $k=1$
    \begin{align}
      (\mu+\lambda \Gamma)\phi_{i,i,1}&=\sum_{l}\phi_{l,i,1}q_{li}+\mu \pi_i.
    \end{align}
    \item $k=\Gamma$
    \begin{align}
     \mu\phi_{i,i^*,\Gamma}=\lambda \Gamma \sum_{l}\phi_{i,l,\Gamma-1}+\sum_{l} \phi_{l,i^*,\Gamma}q_{li}.
    \end{align}
\end{itemize}
Let $\Phi_k=\{\phi_{i,j,k}\}_{i,j\in \mathcal{S}}$ for $1\leq k<\Gamma$, be matrices of size $S \times S$ and let $\Phi_{\Gamma}=\{\phi_{i,i^*,\Gamma}\}_{i\in \mathcal{S}}$ be a column vector. Then, the above equations can be denoted in matrix notation as follows:
\begin{itemize}
    \item $k=1$
    \begin{align}
        (\mu+\lambda \Gamma)\Phi_1=Q^T\Phi_1+\mu \Pi.\label{eqn:mat_1}
    \end{align}
    \item $1\leq k<\Gamma$
    \begin{align}
         (\mu+\lambda \Gamma)\Phi_k=Q^T\Phi_k+\lambda \Gamma \Phi_{k-1}.\label{eqn:mat_k}
    \end{align}
    \item $k=\Gamma$
    \begin{align}
        \mu \Phi_\Gamma=\lambda \Gamma \Phi_{\Gamma-1}\mathbf{1}+Q^T\Phi_\Gamma.\label{eqn:col_K}
    \end{align}
\end{itemize}
Let $\tilde{Q}=\left((\mu+\lambda \Gamma)I-Q^T\right)^{-1}$. From \eqref{eqn:mat_1}, we have,
\begin{align}
    \Phi_1=\mu\tilde{Q}\Pi. \label{eqn:Phi_1}
\end{align}
From \eqref{eqn:Phi_1} and \eqref{eqn:mat_k}, we have for $1<k<\Gamma$,
\begin{align}
    \Phi_k=\mu(\lambda \Gamma)^{k-1}\tilde{Q}^k\Pi. \label{eqn:Phi_k}
\end{align}
Finally, from \eqref{eqn:col_K} and \eqref{eqn:Phi_k} we get,
\begin{align}
    \Phi_\Gamma= \mu(\lambda \Gamma)^{\Gamma-1}\hat{Q}\tilde{Q}^{\Gamma-1}\Pi\mathbf{1}
          =\mu(\lambda \Gamma)^{\Gamma-1}\hat{Q}\tilde{Q}^{\Gamma-1}\bm{\pi}.\label{eqn:col_Phi_K}
\end{align}
where $\hat{Q}=\left(\mu I-Q^T\right)^{-1}$.
Note that since $\bm{\pi}^TQ=\bm{0}$, we have that $\bm{\pi}$ is an eigenvector of $\tilde{Q}$ and $\hat{Q}$ with eigenvalues $\frac{1}{(\mu+\lambda \Gamma)}$ and $\frac{1}{\mu}$, respectively. Hence, \eqref{eqn:col_Phi_K} can be further simplified to the following,
\begin{align}
    \Phi_\Gamma= \frac{(\lambda \Gamma)^{\Gamma-1}}{(\mu+\lambda \Gamma)^{\Gamma-1}}\bm{\pi}.
\end{align}
Now, $\e[\Delta_{\Gamma,\lambda}]$ can be found as follows,
\begin{align}
    \e[\Delta_{\Gamma,\lambda}]&=\sum_{k=1}^{\Gamma-1}\sum_{i}\phi_{i,i,k}+\phi_{i^*,i^*,\Gamma}\\
    &=\sum_{k=1}^{\Gamma-1}\text{tr}\left(\Phi_k\right)+\bm{e}_{i^*}^T\Phi_\Gamma\\
    &=\sum_{k=1}^{\Gamma-1}\text{tr}\left(\Phi_k\right)+\bm{e}_{i^*}^T\frac{(\lambda \Gamma)^{\Gamma-1}}{(\mu+\lambda \Gamma)^{\Gamma-1}}\bm{\pi}\\
    &=\sum_{k=1}^{\Gamma-1}\text{tr}\left(\Phi_k\right)+\frac{(\lambda \Gamma)^{\Gamma-1}}{(\mu+\lambda \Gamma)^{\Gamma-1}}\pi_{i^*}.
\end{align}

\section{Proof of Corollary~\ref{cor:erl_tr}} \label{appen:cor_erl_tr}
Since $Q$ is time reversible, we have that $Q=\Pi^{-\frac{1}{2}}UDU^T\Pi^{\frac{1}{2}}$. This gives us $\tilde{Q}=\Pi^{\frac{1}{2}}U\tilde{D}U^T\Pi^{-\frac{1}{2}}$ where $\tilde{D}$ is a diagonal matrix with $\tilde{D}_{ii}=\frac{1}{d_i+\mu+\lambda\Gamma}$. Therefore, we have $\tilde{Q}^k= \Pi^{\frac{1}{2}}U\tilde{D}^kU^T\Pi^{-\frac{1}{2}}$. Hence, we have,
\begin{align}
    \text{tr}\left(\tilde{Q}^k\Pi\right)&=\text{tr}\left(\Pi^{\frac{1}{2}}U\tilde{D}^kU^T\Pi^{-\frac{1}{2}}\Pi\right)\\
    &=\text{tr}\left(U\tilde{D}^kU^T\Pi\right)\\
    &=\text{tr}\left(\sum_{i=1}^S\frac{1}{(d_i+\mu+\lambda\Gamma)^k}\bm{u}_i\bm{u}_i^T\Pi\right)\\
    &=\sum_{i=1}^S\frac{1}{(d_i+\mu+\lambda\Gamma)^k}\text{tr}\left(\bm{u}_i\bm{u}_i^T\Pi\right)\\
    &=\sum_{i=1}^S\frac{1}{(d_i+\mu+\lambda\Gamma)^k}\sum_{j=1}^S\pi_ju_{i,j}^2\\
    &=\sum_{i=1}^S\frac{a_i}{(d_i+\mu+\lambda\Gamma)^k}.
\end{align}
Then, substituting for $\text{tr}\left(\tilde{Q}^k\Pi\right)$ in \eqref{eqn:erlang_BF} yields,
\begin{align}
    \e[\Delta_{\Gamma,\lambda}]=&\frac{\mu}{\lambda\Gamma}\sum_{k=1}^{\Gamma-1} \sum_{i=1}^Sa_i\left(\frac{\lambda\Gamma}{d_i+\mu+\lambda\Gamma}\right)^k+\frac{(\lambda \Gamma)^{\Gamma-1}}{(\mu+\lambda \Gamma)^{\Gamma-1}}\pi_{i^*}\nonumber\\
    =&\sum_{i=1}^Sa_i\frac{\mu}{\lambda\Gamma}\sum_{k=1}^{\Gamma-1} \left(\frac{\lambda\Gamma}{d_i+\mu+\lambda\Gamma}\right)^k+\frac{(\lambda \Gamma)^{\Gamma-1}}{(\mu+\lambda \Gamma)^{\Gamma-1}}\pi_{i^*}\nonumber\\
    =&\sum_{i=1}^Sa_i\frac{\mu}{d_i+\mu}\left( 1-\left(\frac{\lambda\Gamma}{d_i+\mu+\lambda\Gamma}\right)^{\Gamma-1}\right)
    +\frac{(\lambda \Gamma)^{\Gamma-1}}{(\mu+\lambda \Gamma)^{\Gamma-1}}\pi_{i^*}.
\end{align}

\section{Proof of Theorem~\ref{thrm:mart_v_map}} \label{appen:mart_v_map}
Let $S_i$ be defined as in Section~\ref{sec:exp_est}. Since the sampling process is regenerative, we have 
\begin{align}
    \e[\Delta_{\tau}]=\frac{\e[\int_0^{S_1}\mathds{1}\{X(t)=\hat{X}_{\tau}(t)\}\dd{t}]}{\e[S_1]},
\end{align}
where $X(0)=i$ with probability $\pi_i$. Now, note that for $t<\tau^*$, we have $\hat{X}_M(t)=\hat{X}_{\tau}(t)=X(0)$. Further, $\e[S_1]=\mu$. Therefore, we have
\begin{align}
   &\e\left[\int_0^{S_1}\mathds{1}\{X(t)=\hat{X}_{\tau}(t)\}\dd{t}\bigg\lvert S_1,S_1>\tau^*,X(0)=i\right]\nonumber\\
   &=\int_0^{S_1}\e[\mathds{1}\{X(t)=\hat{X}_{\tau}(t)|X(0)=i\}]\dd{t}\\
   &=\int_0^{\tau^*}\e[\mathds{1}\{X(t)=i\}]\dd{t}+\int_{\tau^*}^{S_1}\e[\mathds{1}\{X(t)=i^*\}]\dd{t}\\
   &=\int_0^{\tau^*}\e[\mathds{1}\{X(t)=i\}]\dd{t}+\int_{\tau^*}^{S_1}P_{ii^*}(t)\dd{t}\\
   &\geq\int_0^{\tau^*}\e[\mathds{1}\{X(t)=i\}]\dd{t}+\int_{\tau^*}^{S_1}P_{ii}(t)\dd{t}\label{eqn:map_ineq}\\
   &=\int_0^{\tau^*}\e[\mathds{1}\{X(t)=i\}]\dd{t}+\int_{\tau^*}^{S_1}\e[\mathds{1}\{X(t)=i\}]\dd{t}\\
   &=\int_0^{S_1}\e[\mathds{1}\{X(t)=\hat{X}_{M}(t)|X(0)=i\}]\dd{t}\\
   &=\e\left[\int_0^{S_1}\mathds{1}\{X(t)=\hat{X}_{M}(t)\}\dd{t}\bigg\lvert S_1,S_1>\tau^*,X(0)=i\right].
\end{align}
The inequality \eqref{eqn:map_ineq} follows from the fact that $i^*$ is the MAP estimator for $X(t)$ when $t>\tau^*$ as a consequence of Lemma~\ref{lem:T}. Further, the interchange of the integral and the expectation is justified by Tonelli's theorem\cite{zygmund}. From the above relation, we have that,
\begin{align}
    \e&\left[\int_0^{S_1}\mathds{1}\{X(t)=\hat{X}_{\tau}(t)\}\dd{t}\bigg\lvert S_1>\tau^*\right]
    \geq \e\left[\int_0^{S_1}\mathds{1}\{X(t)=\hat{X}_{M}(t)\}\dd{t}\bigg\lvert S_1>\tau^*\right]. \label{partial-res1}
\end{align}
Also, since $\hat{X}_M(t)=\hat{X}_{\tau}(t)$ for $t<\tau^*$, we further have that,
\begin{align}
     \e&\left[\int_0^{S_1}\mathds{1}\{X(t)=\hat{X}_{\tau}(t)\}\dd{t}\bigg\lvert S_1\leq\tau^*\right]= \e\left[\int_0^{S_1}\mathds{1}\{X(t)=\hat{X}_{M}(t)\}\dd{t}\bigg\lvert S_1\leq\tau^*\right]. \label{partial-res2}
\end{align}
Combining \eqref{partial-res1} and \eqref{partial-res2} gives the desired result.

\section{Proof of Theorem~\ref{thrm:fresh}} \label{appen:thrm:fresh}
Let $T_{i,m}$ be the time elapsed till we get the next sample starting from state $i$ for the $m$th time. Let $F_{i,m}$ be the proportion of time the estimator was fresh during the time interval $T_{i,m}$. Let $N_i(t)$ be the number of times we took a sample from state $i$ by time $t$ and let $N(t)=\sum_{j=1}^SN_j(t)$. Note that, as $T\to \infty$, $N_i(T)\to \infty$, and hence, so does $N(T)$. Then, the time averaged binary freshness will be given by,
\begin{align}
    \Delta&=\lim_{T\to\infty}\frac{1}{T}\int_0^T\mathds{1}\{X(t)=\hat{X}(t)\}\dd{t}\label{eqn:f1}\\ 
    &=\lim_{N(T)\to\infty} \frac{\sum_{i=1}^S\sum_{m=1}^{N_i(T)}F_{i,m}}{\sum_{i=1}^S\sum_{m=1}^{N_i(T)}T_{i,m}}\label{eqn:f2}\\ 
    &=\lim_{N(T)\to\infty} \frac{\sum_{i=1}^S\frac{N_i(T)}{N(T)}\frac{1}{N_i(T)}\sum_{m=1}^{N_i(T)}F_{i,m}}{\sum_{i=1}^S\frac{N_i(T)}{N(T)}\frac{1}{N_i(T)}\sum_{m=1}^{N_i(T)}T_{i,m}}. \label{eqn:f3}
\end{align}
In \eqref{eqn:f1}, we replaced $\limsup$ with $\lim$ since the limits of the subsequent equations hold, and equality in \eqref{eqn:f2} holds almost surely. Now, as  $N_i(T)\to\infty$, $\frac{1}{N_i(T)}\sum_{m=1}^{N_i(T)}F_{i,m}\to\e[F_i]$ and $\frac{1}{N_i(T)}\sum_{m=1}^{N_i(T)}T_{i,m}\to\frac{1}{\mu}$.
Then, all that remains is to find the  ratios, $\frac{N_i(T)}{N(T)}$, as $N(T)\to\infty$. Let $T_i$ denote the portion of time that the CTMC was in state $i$ during the time $T$. Then, we have that $\lim_{T\to\infty}\frac{T_i}{T}=\pi_i$ and since sampling is a Poisson process, then $\lim_{T\to \infty}\frac{N_i(T)}{T_i}=\mu$. Hence, the limiting ratio can be found as follows,
\begin{align}
    \lim_{N_i(T)\to\infty}\frac{N_i(T)}{N(T)}&=\lim_{T\to \infty}\frac{\frac{N_i(T)}{T_i}\frac{T_i}{T}}{\sum_{j=1}^S\frac{N_j(T)}{T_j}\frac{T_j}{T}}=\frac{\mu\pi_i}{\sum_{j=1}^S\mu\pi_j}=\pi_i.
\end{align}
Plugging this in \eqref{eqn:f3}, along with the fact that $\sum_{j=1}^S\pi_i=1$, yields the desired result.

\section{Proof of Theorem~\ref{thrm:F_i}} \label{appen:thrm:F_i}
Let $Y$ be an exponential random variable with rate $\mu$. Then, $\e[F_i]$ can be found as follows,
\begin{align}
    \e[F_i]&=\e\left[\int_0^{Y}\mathds{1}\{X(t)=\hat{X}(t)\}\dd{t}\Big\lvert X(0)=i\right]\\
           &=\e\left[\int_0^{\infty}\mathds{1}\{X(t)=\hat{X}(t)\}\mathds{1}\{Y>t\}\dd{t}\Big\lvert X(0)=i\right]\label{eqn:Fi_1}\\
           &=\int_0^{\infty}\e\left[\mathds{1}\{X(t)=\hat{X}(t)\}\Big\lvert X(0)=i\right]\e\left[\mathds{1}\{Y>t\}\right]\dd{t}\label{eqn:Fi_2}\\
           &=\int_0^{\infty}P_{i\hat{X}(t)}(t)e^{-\mu t}\dd{t}.
\end{align}
In \eqref{eqn:Fi_2}, the interchange of the expectation and the integral is justified by Tonelli's theorem, and the decomposition of the expectation arises from the fact that $Y$ is independent of both $X(t)$ and $\hat{X}(t)$.

\section{Proof of Theorem~\ref{thrm:fresh_mui}} \label{appen:thrm:fresh_mui}
From \eqref{eqn:f3}, we have that,
\begin{align}
    \tilde{\Delta}&=\lim_{N(T)\to\infty} \frac{\sum_{i=1}^S\frac{N_i(T)}{N(T)}\frac{1}{N_i(T)}\sum_{m=1}^{N_i(T)}F_{i,m}}{\sum_{i=1}^S\frac{N_i(T)}{N(T)}\frac{1}{N_i(T)}\sum_{m=1}^{N_i(T)}T_{i,m}}\\
&=\frac{\sum_{i=1}^S\left(\lim_{N(T)\to\infty} \frac{N_i(T)}{N(T)}\right)\e[\tilde{F}_{i,\mu_i}]}{\sum_{i=1}^S\left(\lim_{N(T)\to\infty} \frac{N_i(T)}{N(T)}\right)\frac{1}{\mu_i}}\label{eqn:f4},
\end{align}
where we have used, as  $N_i(T)\to\infty$, $\frac{1}{N_i(T)}\sum_{m=1}^{N_i(T)}F_{i,m}\to\e[\tilde{F}_{i,\mu_i}]$ and $\frac{1}{N_i(T)}\sum_{m=1}^{N_i(T)}T_{i,m}\to\frac{1}{\mu_i}$. Now, to find the limiting ratios, $\lim_{N(T)\to\infty} \frac{N_i(T)}{N(T)}$, we utilize the two-dimensional CTMC $\tilde{Z}(t)$ constructed for the ME. Note that, whenever the state of $\tilde{Z}(t)$ is in the form $\tilde{Z}(t)=(.,j)$, we will be sampling at the rate $\mu_j$. Let $N_{i,j}(T)$ be the number of times we sampled state $i$  when sampling at rate $\mu_j$. Let $\hat T_{i,j}$ be the proportion of time the chain $\tilde{Z}(t)$ spent on state $(i,j)$ by  time $T$. Since sampling is a Poisson process, we have that $\lim_{T\to\infty}\frac{N_{i,j}(T)}{\hat T_{i,j}}=\mu_j$. Further, we have that $\lim_{T\to\infty}\frac{\hat T_{i,j}}{T}=\psi_{ij}$. Then, we have that,
\begin{align}
    \lim_{N(T)\to\infty} \frac{N_i(T)}{N(T)}&=\lim_{N(T)\to\infty} \frac{\sum_{j\in \mathcal{S}}\frac{N_{i,j}(T)}{\hat T_{i,j}}\frac{\hat T_{i,j}}{T}}{\sum_{k\in \mathcal{S}}\sum_{j\in \mathcal{S}}\frac{N_{k,j}(T)}{\hat T_{k,j}}\frac{\hat T_{k,j}}{T}}\\
    &=\frac{\sum_{j\in \mathcal{S}}\mu_j\psi_{ij}}{\sum_{k\in \mathcal{S}}\sum_{j\in \mathcal{S}}\mu_j\psi_{kj}}\label{eqn:ratio_1}.
\end{align}
To further simplify \eqref{eqn:ratio_1}, we use the global balanced equations of $\tilde{Z}(t)$. Using the fact $\bm\psi^TQ_M=\bm{0}$, we have the following relations for $i\neq j$,
\begin{align}
    \psi_{ii}\mu_i&=\sum_{k\in \mathcal{S}}\psi_{ki}q_{ki}+\sum_{k\in \mathcal{S}}\psi_{ik}\mu_k,\label{eqn:psi_1}\\
    \psi_{ji}\mu_i&=\sum_{k\in \mathcal{S}}\psi_{ki}q_{kj}.\label{eqn:psi_2}
\end{align}
Summing \eqref{eqn:psi_2} across all $j\neq i$, we get,
\begin{align}
    \sum_{j\neq i}\psi_{ji}\mu_i=\sum_{j\neq i}\sum_{k\in \mathcal{S}}\psi_{ki}q_{kj}.\label{eqn:psi_3}
\end{align}
Now, adding \eqref{eqn:psi_3} to \eqref{eqn:psi_1}, we have,
\begin{align}
    \mu_i\sum_{j\in \mathcal{S}}\psi_{ji}&=\sum_{j\in \mathcal{S}}\sum_{k\in \mathcal{S}}\psi_{ki}q_{kj}+\sum_{k\in \mathcal{S}}\psi_{ik}\mu_k\\
    &=\sum_{k\in \mathcal{S}}\psi_{ki}\underbrace{\sum_{j\in \mathcal{S}}q_{kj}}_{=0}+\sum_{k\in \mathcal{S}}\psi_{ik}\mu_k\\
    &=\sum_{k\in \mathcal{S}}\psi_{ik}\mu_k.
\end{align}
Thus, \eqref{eqn:ratio_1} reduces to,
\begin{align}
    \lim_{N(T)\to\infty} \frac{N_i(T)}{N(T)}=\frac{\mu_i\sum_{j\in \mathcal{S}}\psi_{ji}}{\sum_{k\in \mathcal{S}}\mu_k\sum_{j\in \mathcal{S}}\psi_{jk}}
    =\frac{\mu_i\tilde{\pi}_i}{\sum_{k\in \mathcal{S}}\mu_k\tilde{\pi}_k}.\label{eqn:ratio_2}
\end{align}
Plugging \eqref{eqn:ratio_2} into \eqref{eqn:f4}, along with the fact $\sum_{i\in \mathcal{S}}\tilde{\pi}_i=\sum_{i\in \mathcal{S}}\sum_{j\in \mathcal{S}}\psi_{ji}=1$, yields the desired result.

\section{Proof of Theorem~\ref{thrm:semi_simple}} \label{appen:thrm_semi_simple}
Again, from \eqref{eqn:f3}, we have that,
\begin{align}
    \tilde{\Delta}=\lim_{N(T)\to\infty} \frac{\sum_{i=1}^S\frac{N_i(T)}{N(T)}\frac{1}{N_i(T)}\sum_{m=1}^{N_i(T)}F_{i,m}}{\sum_{i=1}^S\frac{N_i(T)}{N(T)}\frac{1}{N_i(T)}\sum_{m=1}^{N_i(T)}T_{i,m}},\label{eqn:semi_f1}
\end{align}
where $N_i(T)\to\infty$, $\frac{1}{N_i(T)}\sum_{m=1}^{N_i(T)}F_{i,m}\to\e[\tilde{F}_{i,\mu_i}]$ and $\frac{1}{N_i(T)}\sum_{m=1}^{N_i(T)}T_{i,m}\to\frac{1}{\mu_i}$ for all $i\neq r$. For state $r$, let us denote by $N_{r,1}(T)$ and $N_{r,2}(T)$, the number of times, starting from state $r$, we used the sampling rate $\mu_{r,1}$ and $\mu_{r,2}$, respectively, to obtain the next sample. Then, we have $N_r(T)=N_{r,1}(T)+N_{r,2}(T)$, with $\lim_{T\to\infty} \frac{N_{r,1}(T)}{N_r(t)}=p$ and $\lim_{T\to\infty} \frac{N_{r,2}(T)}{N_r(t)}=1-p$. Thus, we have that as $N_i(T)\to\infty$, $\frac{1}{N_i(T)}\sum_{m=1}^{N_i(T)}F_{i,m}\to (p\e[F_{r,\mu_{r,1}}]+(1-p)\e[F_{r,\mu_{r,2}}])$ whereas  $\frac{1}{N_i(T)}\sum_{m=1}^{N_i(T)}T_{i,m}\to \frac{p}{\mu_{r,1}}+\frac{(1-p)}{\mu_{r,2}}$. The limiting ratios $\lim_{T\to\infty}\frac{N_{i}(T)}{N(T)}$ only depend on the net sampling rate of each state. Therefore, we have $\lim_{T\to\infty}\frac{N_{i}(T)}{N(T)}=\frac{\mu^{(r)}_i\tilde{\pi}^{(r)}_i}{\sum_{k\in \mathcal{S}}\mu^{(r)}_k\tilde{\pi}^{(r)}_k}$. Substituting the above relations in \eqref{eqn:semi_f1} gives the MBF. The average sampling rate can be obtained by finding the average inter-sampling interval as follows,
\begin{align}
    &\lim_{T\to \infty}\sum_{i=1}^S\frac{N_i(T)}{N(T)}\frac{1}{N_i(T)}\sum_{m=1}^{N_i(T)}T_{i,m}\nonumber\\
    &=\frac{1}{\sum_{k\in \mathcal{S}}\mu^{(r)}_k\tilde{\pi}^{(r)}_k}\left(\mu^{(r)}_r\tilde{\pi}^{(r)}_r\left(\frac{p}{\mu_{r,1}}+\frac{(1-p)}{\mu_{r,2}}\right)+\sum_{i\neq r}\mu_i^{(r)}\tilde{\pi}^{(r)}_i\frac{1}{\mu_i}\right)\\
    &=\frac{1}{\sum_{k\in \mathcal{S}}\mu^{(r)}_k\tilde{\pi}^{(r)}_k}\left(\frac{\tilde{\pi}^{(r)}_r}{\mu_{r,1}\mu_{r,2}}\left(p\mu_{r,1}+(1-p)\mu_{r,2}\right)\left(p\mu_{r,2}+(1-p)\mu_{r,1}\right)+(1-\tilde{\pi}^{(r)}_r)\right).\label{eqn:avg_samp_int}
\end{align}
Taking the reciprocal of the expression in \eqref{eqn:avg_samp_int} yields the average sampling rate.

\section{Proof of Lemma~\ref{lem:continuity}} \label{appen:lem:continuity}
Let $\eta>0$. First, we have,
\begin{align}
    L^\theta(\tilde{\bm\mu}_\theta)=F(\tilde{\bm\mu}_\theta)-\theta J(\tilde{\bm\mu}_\theta)\geq F(\tilde{\bm\mu}_{\theta+\eta})-\theta J(\tilde{\bm\mu}_{\theta+\eta})=L^{\theta+\eta}(\tilde{\bm\mu}_{\theta+\eta})+\eta J(\tilde{\bm\mu}_{\theta+\eta}).\label{eqn:monoton_1}
\end{align}
Therefore, $L^\theta(\tilde{\bm\mu}_\theta)\geq L^{\theta+\eta}(\tilde{\bm\mu}_{\theta+\eta})$. Next, we have,
\begin{align}
     L^{\theta+\eta}(\tilde{\bm\mu}_{\theta+\eta})=F(\tilde{\bm\mu}_{\theta+\eta})-(\theta+\eta) J(\tilde{\bm\mu}_{\theta+\eta})\geq F(\tilde{\bm\mu}_\theta)-(\theta+\eta) J(\tilde{\bm\mu}_\theta)=L^\theta(\tilde{\bm\mu}_\theta)-\eta J(\tilde{\bm\mu}_\theta).
\end{align}
Therefore, we have $\eta J(\tilde{\bm\mu}_\theta)\geq L^\theta(\tilde{\bm\mu}_\theta)-L^{\theta+\eta}(\tilde{\bm\mu}_{\theta+\eta})\geq \eta J(\tilde{\bm\mu}_{\theta+\eta})$. Hence $J(\tilde{\bm\mu}_\theta)\geq J(\tilde{\bm\mu}_{\theta+\eta})$. Finally, from the monotonicity of $J(\tilde{\bm\mu}_\theta)$  and \eqref{eqn:monoton_1}, we have 
$F(\tilde{\bm\mu}_\theta)\geq F(\tilde{\bm\mu}_{\theta+\eta})$. These above relations are inspired from \cite{Ross_MDP, Ross_CSMDP} and have been stated in here since they directly translate to what follows next.

Now, to prove the continuity results, let $F^i(\tilde{\mu}_i)=w_i\e[\Delta^{(i)}]$, which is the weighted MBF of the $i$th CTMC when the sampling rate $\tilde{\mu}_i$ is enforced. Let $L_i^\theta(\tilde{\mu}_i)=F^i(\tilde{\mu}_i)-\theta \tilde{\mu}_i$ and  $\tilde{\mu}_i^\theta$ be its maximizer as defined earlier. Using similar arguments as in the monotonicity proof of $L^\theta(\tilde{\bm\mu}_\theta)$, $F(\tilde{\bm\mu}_\theta)$ and $J(\tilde{\bm\mu}_\theta)$, it follows that $L_i^\theta(\tilde{\mu}_i^\theta)$, $F^i(\tilde{\mu}_i^\theta)$ and $\tilde{\mu}_i^\theta$ are also monotonically decreasing functions with respect to $\theta$. Further, we have that,
\begin{align}
    L_i^\theta(\tilde{\mu}_i^\theta)-L_i^{\theta+\eta}(\tilde{\mu}_i^{\theta+\eta})\leq L_i^\theta(\tilde{\mu}_i^\theta)-L_i^{\theta+\eta}(\tilde{\mu}_i^{\theta})=\eta\tilde{\mu}_i^{\theta}\leq \eta\tilde{\rho}.
\end{align}
Therefore, we have that $L_i^\theta(\tilde{\mu}_i^\theta)$ and thereby $L^\theta(\tilde{\bm\mu}_\theta)$ are continuous with respect to $\theta$. Now, suppose in the event that there are multiple maximizers for $L_i^\theta(\tilde{\mu}_i)$, we always choose  $\tilde{\mu}_i^\theta$ as the smallest maximizer. In this case, we will show that $\tilde{\mu}_i^\theta$ is right-continuous with respect to $\theta$. First, assume $\tilde{\mu}_i^\theta$ is not right-continuous. Then, we have that there exists some $\theta$ and $\epsilon>0$ such that $\forall \eta>0$, there is a $\theta'\in [\theta,\theta+\eta)$ such that $\tilde{\mu}_i^\theta-\tilde{\mu}_i^{\theta'}\geq\epsilon$. This allows us to choose a decreasing sequence of $\theta_n$ such that $\theta_n\to \theta$ and for each $\theta_n$ we have $\tilde{\mu}_i^\theta\geq \tilde{\mu}_i^{\theta_n}+\epsilon$. Since $\tilde{\mu}_i^{\theta_n}$ is a decreasing sequence, its limit exists and let us denote it by $\bar\mu_i=\lim_{n\to\infty}\tilde{\mu}_i^{\theta_n}$. Since $L_i^\theta(\tilde{\mu}_i^\theta)$ is continuous with respect to $\theta$, we have $\lim_{n\to \infty}L_i^{\theta_n}(\tilde{\mu}_i^{\theta_n})=L_i^\theta(\tilde{\mu}_i^\theta)$. Further, since $F(\tilde{\mu}_i)$ is continuous with respect to $\tilde{\mu}_i$, we have that $\lim_{n\to \infty}L_i^{\theta_n}(\tilde{\mu}_i^{\theta_n})=F(\bar\mu_i)-\theta\bar\mu_i$. Therefore, $\bar\mu_i$ is also a maximizer for $L_i^\theta(\tilde{\mu}_i)$ with $\bar\mu_i+\epsilon\leq\tilde{\mu}_i^\theta$. This is a contradiction since $\tilde{\mu}_i^\theta$  is the smallest maximizer. Therefore, $\tilde{\mu}_i^\theta$ is right-continuous with respect to $\theta$. Similarly, we can show that, if we always choose the largest maximizer, then $\tilde{\mu}_i^\theta$ is left-continuous with respect to $\theta$. Combining these two facts gives us that if $\tilde{\mu}_i^\theta$ is unique for all $\theta$, then $\tilde{\mu}_i^\theta$ is continuous, and hence, so is $J(\tilde{\bm\mu}^\theta)$.

\bibliographystyle{unsrt}
\bibliography{refs}
\end{document}